\documentclass[notitlepage,aps,pra,reprint,twocolumn,superscriptaddress]{revtex4-1}
\usepackage{graphicx}
\usepackage{amsmath}
\usepackage{amssymb}
\usepackage{comment}
\usepackage[colorlinks, allcolors=blue]{hyperref}
\usepackage[all]{hypcap}
\usepackage[mathlines]{lineno}
\usepackage{braket}
\usepackage{wrapfig}
\usepackage{lipsum}
\usepackage{ulem}
\usepackage{physics}

\newcommand{\pref}[2]{\hyperref[#1]{\ref{#1}(#2)}}
\newcommand{\preff}[2]{\hyperref[#1]{\ref{#1 b}#2}}
\newcommand{\eqpref}[1]{\hyperref[#1]{(\ref{#1})}}
\newcommand{\red}[1]{\textcolor{black}{#1}}
\newcommand{\pt}{$\mathcal{PT}$}

\newcommand{\squig}{{\raise.17ex\hbox{$\scriptstyle\sim$}}}

\begin{document}
\title{Coexistence of stable and unstable population dynamics in a nonlinear non-Hermitian mechanical dimer}

\author{Enrico Martello}
\affiliation{School of Physics and Astronomy, University of Birmingham, Edgbaston, Birmingham B15 2TT, United Kingdom}
\author{Yaashnaa Singhal}
\affiliation{Department of Physics, University of Illinois at Urbana-Champaign, Urbana, Illinois 61801-3080, USA}
\author{Bryce Gadway}
\affiliation{Department of Physics, University of Illinois at Urbana-Champaign, Urbana, Illinois 61801-3080, USA}
\author{Tomoki Ozawa}
\affiliation{Advanced Institute for Materials Research (WPI-AIMR), Tohoku University, Sendai 980-8577, Japan}
\author{Hannah M. Price}
\affiliation{School of Physics and Astronomy, University of Birmingham, Edgbaston, Birmingham B15 2TT, United Kingdom}
\date{\today}

\begin{abstract}
Non-Hermitian two-site ``dimers" serve as minimal models in which to explore the interplay of gain and loss in dynamical systems. In this paper, we experimentally and theoretically investigate the dynamics of non-Hermitian dimer models with non-reciprocal hoppings between the two sites. We investigate two types of non-Hermitian couplings; one is when asymmetric hoppings are externally introduced, and the other is when the non-reciprocal hoppings depend on the population imbalance between the two sites, thus introducing the non-Hermiticity in a dynamical manner. We engineer the models in our synthetic mechanical set-up comprised of two classical harmonic oscillators coupled by measurement-based feedback. For fixed non-reciprocal hoppings, we observe that, when the strength of these hoppings is increased, there is an expected transition from a  \pt-symmetric regime, where oscillations in the population are stable and bounded, to a \pt-broken regime, where the oscillations are unstable and the population grows/decays exponentially. However, when the non-Hermiticity is dynamically introduced, we also find a third intermediate regime in which these two behaviors coexist, meaning that we can tune from stable to unstable population dynamics by simply changing the initial phase difference between the two sites. As we explain, this behavior can be understood by theoretically exploring the emergent fixed points of a related dimer model in which the non-reciprocal hoppings depend on the {\it normalized} population imbalance. Our study opens the way for the future exploration of non-Hermitian dynamics and exotic lattice models in synthetic mechanical networks.
\end{abstract}

\maketitle

\section{Introduction}
In recent decades, systems described by non-Hermitian Hamiltonians have become a topic of great interest since gain and loss can lead to many intriguing effects, including parity-time (\pt) symmetry and real energy spectra~\cite{Bender,Bender07,Longhi,El-Ganainy,guo2009,Bittner,Schindler}, exceptional points~\cite{dembowski2001,dembowski2004,gao2015observation,Nori-NatMats-Review}, non-Hermitian geometrical phases~\cite{garrison1988complex, dattoli1990geometrical,  Keck_2003, liang2013, mondragon1996berry, berry2003optical,berry2004physics, nesterov2008complex,singhal2022measuring} and new types of topological phenomena~\cite{Lee2016,shen2018,xiao2020non,yao2018,silveirinha2019topological,ashida2020,Coulais2021, bergholtz2021}. This interest has also been driven by recent experimental developments leading to non-Hermitian physics being simulated across a wide-range of platforms such as photonics~\cite{guo2009,Nori-NatMats-Review,feng2017non,el2019dawn,zhao2019non,price2022roadmap}, ultracold gases~\cite{Bo-NonRecip,li2019observation,liang2022dynamic}
and mechanical metamaterials~\cite{Coulais2017,Brandenbourger2019,scheibner2020,scheibner2020odd,zhou2020non,ghatak2020observation,singhal2022measuring}.

Within this field, significant effort has been devoted to the study of so-called ``dimer" models, in which two sites are coupled together in a (non-Hermitian) Hamiltonian that can be represented generally as a $2 \times 2$ matrix. Such models are useful as they serve as minimal systems in which to understand (often analytically) the effects of non-Hermitian terms on dynamical behavior~\cite{graefe2008,graefe2008mean,graefe2010,graefe2012}. \red{Some of the interesting dynamical effects that have been investigated in nonlinear dimer systems include sensitivity to input power~\cite{alexeeva2014actively,Schindler}, non-reciprocal dynamics and directed transport~\cite{kominis2016asymmetric,ramezani2010unidirectional}, and confinement in phase space~\cite{barashenkov2015dimer}.}
Many of the previous works focused on the dimer models where the non-Hermiticity is introduced and controlled through on-site loss and gain terms (see, \textit{e.g.},~\cite{guo2009,Bittner,Schindler,Vazquez,graefe2008,graefe2008mean,graefe2010,graefe2012,zhiyenbayev2019enhanced,valagiannopoulos2022multistability} and references therein), that is, through the diagonal components in the Hamiltonian.

In this paper, we instead explore dynamics of dimer models in which the non-Hermitian effects are introduced by making the hopping between the two sites asymmetric~\cite{wu2022interplay}. Such systems are inspired by the Hatano-Nelson (HN) model~\cite{hatano96,hatano97,hatano98}, which is a one-dimensional lattice model with asymmetric hoppings between the sites exhibiting the so-called non-Hermitian skin effect~\cite{bergholtz2021,Lee2016,Gong}. Such asymmetric, or non-reciprocal, hoppings are more difficult to experimentally implement than on-site gain and loss. However, there has been significant recent progress in realizing such asymmetric hoppings using setups based on optical systems~\cite{peng2014parity,ruesink2016nonreciprocity,fang2017generalized,weidemann2020topological}, electrical circuits~\cite{ezawa2019non,ezawa2019electric,liu2021non} and synthetic mechanical metamaterials~\cite{MechMeta, singhal2022measuring}, opening up a possibility to experimentally study such non-Hermitian models. 

In this paper, we engineer non-Hermitian dimer models by taking advantage of the flexibility of a mechanical set-up consisting of two coupled harmonic oscillators with measurement-based feedback~\cite{MechMeta}. As we have previously demonstrated, this approach can be used to simulate near-arbitrary mean-field lattice Hamiltonians, with controllable on-site gain and loss, non-reciprocal couplings and (exotic) synthetic nonlinearities amongst other effects~\cite{MechMeta, singhal2022measuring}. In addition to its tunability, a key advantage of this set-up is that it provides full access to the dynamics, allowing us to observe the evolution of the system in real time. 

Here, we exploit this mechanical set-up to investigate the dynamics of HN dimer models. \red{We first discuss the linear HN dimer in which the asymmetric hopping is externally fixed~\cite{wu2022interplay}. This linear model is mathematically equivalent to the aforementioned dimer models with on-site gain and loss, with the result that the trajectories of dynamics in the \pt-symmetric regime show closed stable orbits, while the behavior in the \pt-broken regime is unstable, with the population exponentially exploding or decaying.} Building on the understanding of the linear model, we then explore a nonlinear version of the Hatano-Nelson dimer in which the asymmetric coupling is induced by population imbalance between the two sites. For this model, we again find both a stable regime at low coupling strengths, in which population oscillations are bounded, analogous to the \pt-symmetric regime described above, and an unstable regime at high coupling strengths, in which the population grows/decays, similar to the \pt-broken regime. However, we also find
a new regime at intermediate coupling strengths, where the two types of behavior coexist, allowing us to tune the population dynamics from a stable oscillation to an unstable divergence by simply tuning the initial phase difference between the oscillators. As we discuss, this behavior can be understood by studying the fixed points of a variant of the nonlinear HN dimer model, which we call the instantaneous HN dimer model, in which the non-reciprocal coupling depends on the {\it normalized} population difference between the two sites as introduced below. Our work lays the foundation for exploring the dynamics of more exotic lattice Hamiltonians with non-Hermiticity and mean-field interactions.

The paper is structured as follows: In Sec.~\ref{sec:dimers} we introduce and define the models we explore.
We introduce the linear HN dimer model, the nonlinear HN dimer model, and the instantaneous HN dimer model that helps elucidate the dynamics of the original nonlinear HN dimer model.
In Sec.~\ref{sec:experimental set-up} we describe the experimental approach and set-up, and the details and parameters chosen.
In Sec.~\ref{sec:HN_dimer}, we analyze the linear HN dimer model. We first analytically study the dynamical behavior of the model, and then we compare the results with numerically and experimentally obtained dynamics.
In Sec.~\ref{sec:HN_inst}, we analyze the dynamical behavior of the instantaneous HN dimer model, and we discuss the emergence of multiple fixed points and the structure of the transition between the weakly and strongly interacting regimes. 
In Sec.~\ref{sec:experimental model}, we finally study the nonlinear HN dimer model. We give qualitative explanations of the phase diagram using results obtained in previous sections, and we discuss the coexistence of different phases in the dynamics from both numerical and experimental approaches. 
Finally, in Sec.~\ref{sec:conc}, we draw conclusions and discuss the outlook for this work.

\section{Models}
\label{sec:dimers}

We experimentally realize two types of non-Hermitian dimer models. The first is the {\it linear} Hatano-Nelson (HN) dimer model, in which Hermiticity is broken by externally-tuneable non-reciprocal couplings between the two sites. The second is the {\it nonlinear} HN dimer model, in which the non-reciprocal couplings are instead induced by the population imbalance between the two sites, and hence evolve dynamically, depending on the interparticle interaction strength. To obtain an analytical understanding of the nonlinear HN dimer model, we also theoretically introduce a variant of the nonlinear HN dimer model in which the non-Hermiticity only depends on the {\it normalized} population imbalance between the two sites; such a model describes the dynamics of the nonlinear HN dimer model for a short period of time, and thus we call it an {\it instantaneous} HN dimer model. We shall now introduce the linear, nonlinear, and instantaneous HN dimer model in turn. 

\subsection{Linear Hatano-Nelson Dimer Model}

\begin{figure}
    \centering
    \includegraphics[width=.48\columnwidth]{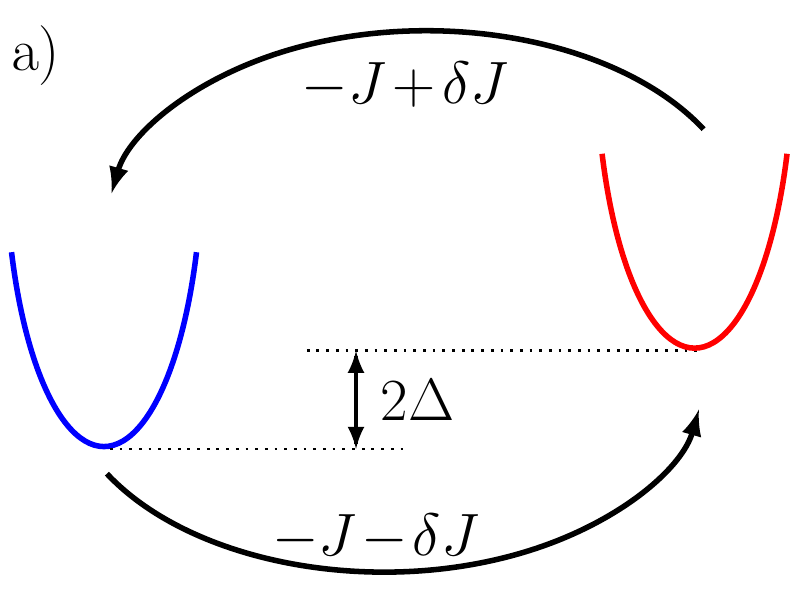} \;%
    \includegraphics[width=.48\columnwidth]{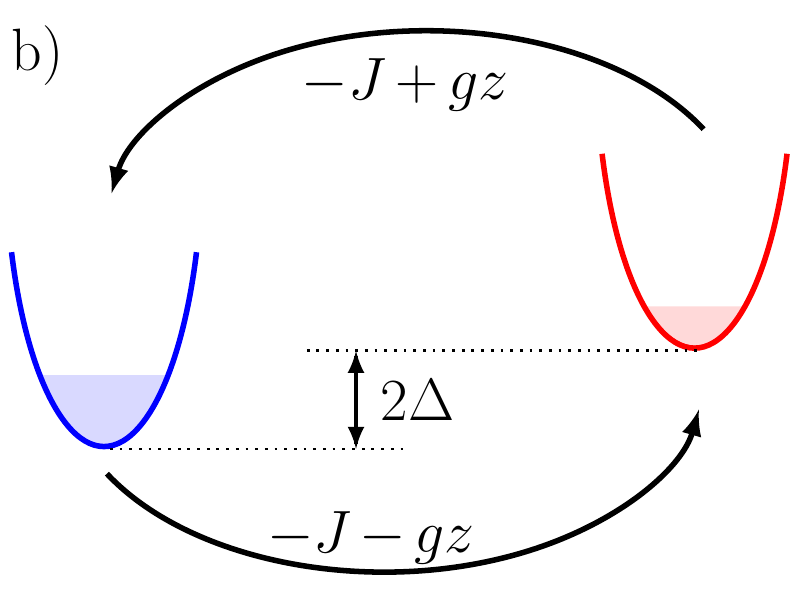} \\ %
    \caption{\textit{a)}~Sketch of the Hatano-Nelson dimer as described in Eq.~\ref{eq:Schroedinger zind} where $\delta J$ is the hopping asymmetry parameter. \textit{b)} Sketch of the dimer with population-dependent hopping asymmetry as described in Eq.~\ref{eq:Schroedinger}, where $z$ describes the population imbalance between the two sites and $g$ is a control parameter.
    }
    \label{fig:model cartoon}
\end{figure}

The linear HN dimer model is a model in which two sites are coupled by non-reciprocal hopping amplitudes~\cite{li2020critical,liu2020non,bergholtz2021,wu2022interplay}. By writing the complex-valued wave-function of the two sites as $\vb* \alpha = (\alpha_1, \alpha_2)^{\text T}$, the dynamics of the linear HN dimer is described by  
\begin{eqnarray}
\label{eq:Schroedinger zind}
 i\dot{\vb* \alpha} = 
\mqty( \omega -  \Delta & -J-\delta J \\ -J+\delta J & \omega+\Delta) \vb* \alpha,
\end{eqnarray}
where $\omega$ is an overall energy offset and  $\Delta$ determines the on-site energy difference between the two sites (with $\hbar=1$), as shown in Fig.~\ref{fig:model cartoon}~(a).
Without loss of generality, from now on we will consider $J$, $\Delta$ and $\delta J$ to be non-negative real values. The coupling between the sites is split into the reciprocal part of the hopping amplitude $-J$ and the non-reciprocal part $ \pm \delta J$, the latter being responsible for breaking Hermiticity. This model is the two-site version of the famous Hatano-Nelson model for a one-dimensional chain~\cite{hatano96,hatano97,hatano98} in which all the nearest-neighbor inter-site couplings take the form as in Eq.~\ref{eq:Schroedinger zind}. 

\red{As mentioned above, this model is mathematically equivalent to the more commonly studied non-Hermitian dimer model with reciprocal coupling and on-site gain and loss, c.f. \textit{e.g.} Refs.~\cite{graefe2008,graefe2008mean,graefe2010,graefe2012}. To see this, the Hamiltonian (i.e.~the two-by-two matrix in Eq.~\ref{eq:Schroedinger zind}) can be expressed with Pauli matrices as: $\mathcal H \!=\! -\Delta \,\sigma_z -J\,\sigma_x -i \delta J\, \sigma_y$, up to an overall energy shift. A suitable unitary transformation can bring this into the form $\mathcal H \!=\! \sqrt{ \Delta^2+J^2}\sigma_x- i \delta J \sigma_z$, which describes a reciprocal hopping of strength $\sqrt{\Delta^2+J^2}$ and on-site imaginary terms of magnitude $\delta J$ with opposite signs corresponding to the gain and loss; the known behavior of this model can therefore be used to infer that of the linear HN model. We will instead analyze the linear HN model directly in Sec.~\ref{sec:HN_dimer} in order to lay the groundwork for the rest of the paper. }

\subsection{Nonlinear Hatano-Nelson Dimer Model}

In the linear HN dimer model, the non-Hermiticity was included via a constant off-diagonal contribution $\pm\delta J$. In the nonlinear HN dimer model, the non-Hermiticity is introduced dynamically via the population imbalance between the two sites. The nonlinear HN dimer model is defined by the following equation of motion:
\begin{eqnarray}
\label{eq:Schroedinger}
 i\dot{\vb* \alpha} = 
\mqty( \omega -  \Delta & -J-gz \\ -J+gz & \omega+\Delta) \vb* \alpha,
\end{eqnarray}
in which the non-reciprocal part of the hopping amplitude is now set by $\pm g z$, where $g$ is the interaction strength and $z\equiv |\alpha_1|^2 - |\alpha_2|^2 $ represents the population imbalance between the two sites. A sketch of this model is made in Fig.~\ref{fig:model cartoon}~(b). As $z$ depends on the complex wave-function, this is no longer a linear model and so does not have linear eigenstates and eigenenergies similar to the linear HN dimer. Understanding the rich behavior of this model is a goal of the current paper. Because of the nonlinearity, it is not possible to obtain a good analytical understanding of the model. In order to approximately understand the dynamics of this model for a short period of time, we introduce the following instantaneous HN dimer model. 

\subsection{Instantaneous Hatano-Nelson Dimer model}
The nonlinearity in the nonlinear HN dimer model was in the term $\pm gz$ in the off-diagonal terms. We can rewrite this term as $\pm gn (z/n)$, where $n$ is the norm of the wavefunction. In non-Hermitian models, the norm $n$ generally depends on time, and so do $gn$ and $z/n$. However, as we discuss in more detail in Sec.~\ref{sec:HN_inst}, the model
\begin{eqnarray}
\label{eq:Schroedinger inst}
 i\dot{\vb* \alpha} = 
\mqty( \omega -  \Delta & -J-\bar{g}z/n \\ -J+ \bar{g}z/n & \omega+\Delta) \vb* \alpha,
\end{eqnarray}
which is obtained by replacing $gn$ in the nonlinear HN dimer model by $\bar{g}$, allows a description of the dynamics without complication arising from the time-dependence of $n$ if $\bar{g}$ is taken to be a constant value. Physically, we can view the model as describing the dynamics of the nonlinear HN dimer model at a time close to $t_0$ if we set $\bar{g} = gn(t_0)$. We thus call the model the instantaneous HN dimer model.

Before moving on to describe the dynamical features of the linear, nonlinear, and instantaneous HN dimer models, we briefly explain how the models can be experimentally realized in our system of mechanical oscillators.
 
\section{Experimental Set-up}
\label{sec:experimental set-up}

Our experimental setup is described in detail in Ref.~\cite{MechMeta}, and it consists of two almost identical oscillators, whose acceleration $a(t)$ and its numerical derivative, the jerk $j(t) =  d a(t)/ dt$, are measured in real time. These measurements can be used to produce a real-time feedback force on each oscillator that allows us to synthetically couple the two oscillators. This procedure enables us to effectively generate a wide variety of two-level tight-binding Hamiltonians. Within the rotating-wave approximation, the \red{mean-field Schr\"odinger equations of motion} for a generic tight-binding Hamiltonian can be mapped onto Newton's equations for the harmonic oscillators. Hence, we can engineer a combination of self and cross feedback~\cite{MechMeta}: the former is responsible for local on-site terms (\textit{e.g.} site-dependent potential energy shifts, site-dependent gain or loss, and on-site nonlinear interactions), while the latter introduces off-site terms that allow energy to hop from site to site (\textit{e.g.}, complex hopping, non-reciprocal coupling, and even density-dependent hopping).

\begin{figure}
\includegraphics[width=.5\columnwidth]{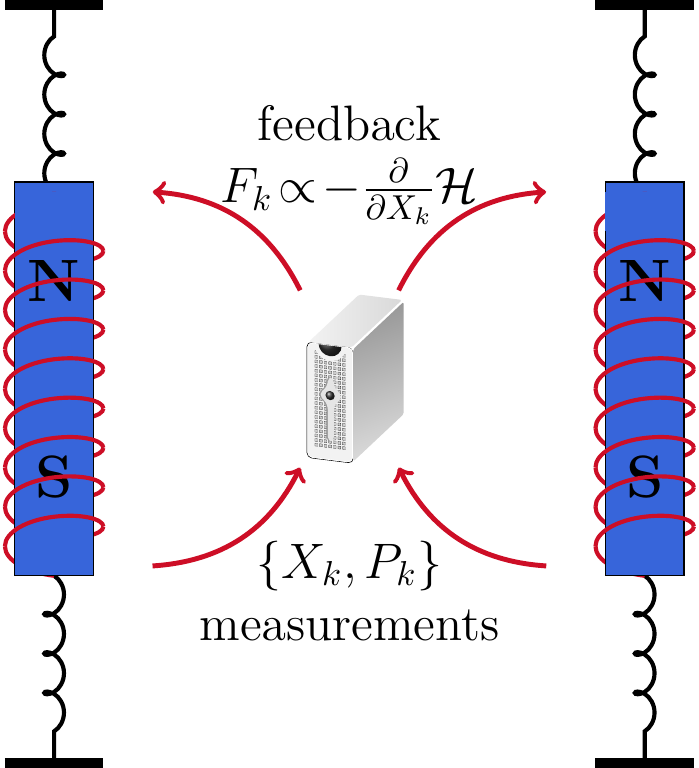}
\caption{Sketch of the experimental apparatus used for implementing the described models. We perform real-time measurements of the effective position $X_k$ and momentum $P_k$ variables of our oscillators (labeled by index $k$), which are used to compute real-time feedback forces that are applied magnetically through a voltage-controlled current. The form of the feedback forces used to implement a desired Hamiltonian $\mathcal{H}$ is given simply by the relationship $F_k \propto -\partial \mathcal{H} / \partial X_k$.
}
\label{fig:exptsystem}
\end{figure}

Experimentally, feedback forces are implemented by using real-time voltage output signals to control the currents through gradient solenoids surrounding each oscillator. These control signals result in magnetic forces on the oscillators, each of which features an embedded dipole magnet. The real-time voltage measurement signals relating to acceleration and jerk are normalized to a common scale. Given the simple harmonic nature of our oscillators, these normalized measurement signals $X_k$ and $P_k$, where $k$ is the oscillator index, serve as direct proxies for the oscillators' positions and momenta, $x_k$ and $p_k$. As depicted in Fig.~\ref{fig:exptsystem}, a desired Hamiltonian $\mathcal{H}$ is implemented by feeding back on the oscillators with forces that are true to Hamilton's equation ($dp_k / dt = -\partial \mathcal{H} / \partial x_k$). Hence, feedback forces are of the form $F_k \propto -\partial \mathcal{H} / \partial X_k$. Naturally, site energy shifts are implemented by forces of the form $F_k \propto X_k$, while gain and loss terms are implemented through feedback of the form $F_k \propto P_k$, and so on~\cite{MechMeta}.
Owing to our co-normalization of the $X_k$ and $P_k$ variables, the \textit{relative} magnitude of all the linear terms in our experimentally implemented Hamiltonians is defined simply by the ratio of the applied feedback coefficients.
The \textit{absolute} calibration of our feedback forces (\textit{i.e.}, how the control voltage signals generated by our measurement-and-feedback system relate to the actual mHz-scale terms of the implemented Hamiltonian) is performed by investigating the frequency shift of the individual oscillators based on self-feedback forces ($F_k \propto X_k$), as detailed in Ref.~\cite{MechMeta}.

To simulate the dynamics as in Eq.~\ref{eq:Schroedinger zind}, we introduce a combination of self-feedback (to cancel natural loss terms and to shift the site energies) and linear cross-feedback (to introduce hopping between sites). To capture nonlinear terms, such as the population-dependent hopping contributions of Eq.~\ref{eq:Schroedinger}, we introduce feedback forces of the form $F_{1(2)} \propto g (X_1^2 + P_1^2 - X_2^2 - P_2^2) X_{2(1)} / (X_{1,i}^2 + P_{1,i}^2 + X_{2,i}^2 + P_{1,i}^2)$, where $X_{k,i}$ and $P_{k,i}$ are the initial values taken by the effective position and momentum variables. Through this normalization, these nonlinear terms are governed by the same absolute calibration as the linear terms.

In addition to allowing for implementation of the desired linear and nonlinear HN dimer models, our control over these applied forces enables us to set the initial state of the oscillators for each experiment. Starting with oscillators nominally at rest, we sinusoidally drive the two oscillators at their common resonance frequency of $\sim 3.05$~Hz for several seconds. By controlling the relative strength and phase of these two sinusoidal ``initialization'' drives, we control the initial amplitudes and phases of the two oscillators (or correspondingly, the initial complex-valued wave-function $\vb* \alpha$).

\section{Linear HN Dimer Model}
\label{sec:HN_dimer}
With the experimental setup we just described, we study the dynamical behavior of the linear HN dimer model. To understand the dynamics, we employ two descriptions, one is in the ``phase space"
of the population imbalance and the phase difference between the two sites, and the other is the Bloch sphere representation. 
\red{Although these dynamics can already be obtained either by looking at the eigenstates of the Hamiltonian or through the mathematical equivalence between the linear HN dimer model and the on-site gain-loss model mentioned in Sec.~\ref{sec:dimers}, the description in terms of the phase space and the Bloch sphere is worth discussing here as it gives us an intuitive picture for understanding, and will also lay the groundwork for later Sections to understand the dynamics of the nonlinear and instantaneous HN dimer models.}

\subsection{Dynamical Equations for the linear HN dimer}

We rewrite here the equation of motion for the linear HN dimer model, Eq.~\ref{eq:Schroedinger zind},
\begin{eqnarray}
 i\frac{\partial }{\partial t} \left(\begin{array}{c} \alpha_1 \\ \alpha_2\end{array} \right) = 
\mqty( \omega -  \Delta & -J-\delta J \\ -J+\delta J & \omega+\Delta) \left(\begin{array}{c} \alpha_1 \\ \alpha_2 \end{array}  \right).
\end{eqnarray}
In standard Hermitian quantum mechanics, the overall normalization of the wavefunction is not an observable, and we can assume the wavefunction to be normalized with the overall phase left as a gauge degree of freedom. However, in non-Hermitian quantum mechanics, time evolution is non-unitary in general and the change of norm of the wavefunction over time plays an important role, describing phenomena such as decay and lasing. In our classical experimental setup, all the information of the wavefunction, $\alpha_1$ and $\alpha_2$, can in principle be measured. Following the practice of non-Hermitian quantum mechanics, we consider the norm of the wavefunction to be an observable, but we choose to not consider the overall phase to be significant.

In describing the dynamics of two-site (two-level) systems, it is both convenient and conventional to recast the dynamical equations either in terms of ``phase space" dynamics~\cite{raghavan} or in terms of the Bloch vector~\cite{graefe2008mean,graefe2010}. As we shall use both pictures interchangeably, we now briefly introduce each in turn. 

In the phase-space picture, we look at the dynamics in the space of phase difference between two sites, $\varphi \equiv \mathrm{arg}(\alpha_1/\alpha_2) \in (-\pi,\pi]$, and the normalized population imbalance between two sites $\tilde z\equiv (|\alpha_1|^2 -|\alpha_2|^2)/n$, where $n\equiv |\alpha_1|^2 +|\alpha_2|^2 $ is the total population. For two-site Hermitian models, the dynamics is fully characterized in the space of $\{ \tilde{z}, \varphi \}$. In non-Hermitian models, however, the overall norm of the wavefunction can change and have significance, so the full dynamics is characterized in the space of $\{ n,\tilde{z},\varphi \}$.
Equations of motion for $\{ n,\tilde{z},\varphi \}$ can be obtained from Eq.~\ref{eq:Schroedinger zind} in a straightforward manner, and they are
\begin{align}
\dot n &=- 2\, \delta J\, n\, \sqrt{1-\tilde z^2} \sin \varphi, \label{eq:n dot zind} \\
\dot {\tilde z} &= 2 ( \tilde z\, \delta J - J) \sqrt{1-\tilde z^2} \,\sin \varphi, \label{eq:z dot zind}\\
\dot \varphi &= -2\, \Delta + 2 \,\frac{J \tilde z - \delta J}{\sqrt{1-\tilde z^2}} \;\cos\varphi. \label{eq:phi dot zind}
\end{align}
We see that the time evolutions of the variables $\tilde{z}$ and $\varphi$ are closed by themselves and do not depend on the total population $n$. As a result, we can consider the dynamics in the restricted 2D phase space $\{ \tilde{z}, \varphi \}$; the dynamics of $n$ is separately determined from the information of $\{ \tilde{z}, \varphi \}$~\cite{graefe2008mean,graefe2010}. Most importantly, when considering the dynamics of $\{ \tilde{z}, \varphi \}$, we do not need to worry about the fact that the total population $n$ can change in time, and thus we can analyze the dynamics in $\{ \tilde{z}, \varphi \}$ in a way analogous to the Hermitian dimer models.

Secondly, we can alternatively visualize the dynamics by means of the normalized Bloch vector, $\vb s  = ( s_x, s_y, s_z)^{\text T}$, whose components are defined as:
\begin{align}
 s_x &\equiv \frac 1{2\,n} \qty(\alpha_1^*\alpha_2 + \alpha_1\alpha_2^*),\label{eq:x comp}\\
 s_y &\equiv \frac 1{2i\,n} \qty(\alpha_1^*\alpha_2 - \alpha_1\alpha_2^*),\label{eq:y comp}\\
 s_z &\equiv \frac 1{2\,n} \qty(|\alpha_1|^2 -|\alpha_2|^2) = \frac 1{2} \tilde z,\label{eq:z comp}
\end{align}
subject to the condition $ |\vb s|^2 = 1/4$, implying that the head of the Bloch vector always lies on the surface of a sphere of radius 1/2 called the Bloch sphere~\cite{graefe2008mean,graefe2010}. As the total population $n$ also changes, the full dynamics is described by the four variables $\{ n, s_x, s_y, s_z\}$.
It can then be shown that~\cite{wu2022interplay}:
\begin{align}
 \dot n \;&=  -4\,\delta J\,n\,s_y \label{eq:fourth zind}\\
 \dot s_x  &= 2 \Delta s_y + 4\,\delta J\,s_x s_y \label{eq:first zind}\\
 \dot s_y &= -2\Delta s_x  + 2 J s_z - \delta J + 4\,\delta J\,s_y^2 \label{eq:second zind}\\
 \dot s_z &= -2J s_y + 4\,\delta J\,s_y s_z, \label{eq:third zind}
\end{align}
where again the time dependence of the Bloch vector $(s_x, s_y, s_z)$ does not depend on the total population $n$, and thus, upon considering the dynamics of the Bloch vector $(s_x, s_y, s_z)$, we do not need to worry about the time dependence of $n$.
We note that since a vector $( s_x, s_y, s_z )$ lies on the surface of a two-dimensional sphere, the dimension of the space in which the variables $\{ n, s_x, s_y, s_z\}$ move is three-dimensional, agreeing with the three-dimensional description in the phase space $\{n, \tilde{z}, \varphi \}$.
Naturally, the phase-space and the Bloch-vector equations are fully equivalent, as can be shown by noting that $\varphi = \arctan({s_y}/{s_x})$ and $ \tilde z = 2  s_z$. 

\subsection{Dynamics of the linear HN dimer}

\red{Linear non-Hermitian dimer models exhibit a \pt-symmetry breaking transition, in which the eigenstates of the Hamiltonian coaelesce at an exceptional point~\cite{Bender,Bender07,heiss} as the strength of the non-Hermitian terms is increased. The energy eigenvalues, $E_{\pm}$, of the Hamiltonian  are~\cite{Bender07,heiss}:
\begin{eqnarray}
\label{eq:en}
E_{\pm}   = \omega   \pm\sqrt{J ^ 2 +\Delta^2 - \delta J^2}. 
\end{eqnarray}
%
The energies are purely real when $J ^ 2 +\Delta^2 > \delta J^2$, corresponding to a weakly non-Hermitian regime called the \pt-symmetric regime. When $J ^ 2 +\Delta^2 = \delta J^2$ the two eigenvalues become degenerate, at which point the eigenvectors also coalesce, yielding the exceptional point in the parameter space. When $J ^ 2 +\Delta^2 < \delta J^2$, the energies acquire a nonzero imaginary part, which is called the \pt-broken regime. The appearance of the imaginary part of eigenenergies indicates that the norm of the wavefunction will not be conserved over time, and the population exponentially grows or decays~\cite{Silberstein,li2020critical,liu2020non,bergholtz2021,wu2022interplay}. In the following, we first review the \pt-symmetric region $J ^ 2 +\Delta^2 > \delta J^2$ where the dynamics is described by oscillation around two fixed points. We then discuss the \pt-broken regime $J ^ 2 +\Delta^2 < \delta J^2$ where the fixed points turn into a source and sink of dynamics. }

\subsubsection{\pt-symmetric Regime}

In the \pt-symmetric regime, $J ^ 2 +\Delta^2 > \delta J^2$, where the eigenvalues of the Hamiltonian are real, the dynamics consist of Rabi oscillations between the two sites, which can be seen as closed orbits in phase-space and on the Bloch sphere~\cite{graefe2008mean,graefe2010}. The population can also be biased toward one of the two sites; the emergence of population imbalance comes from non-Hermiticity, and its mechanism is different from the self-trapping known in interacting Hermitian Josephson dimers.

The dynamics can be understood through the {\it fixed points} of motion, which are obtained by setting the time derivative of the variables to zero. In the non-Hermitian models that we analyze in this paper, we find it convenient and useful to look for fixed points of variables other than $n$, leaving the possibility for $n$ to change in time. In what follows, when we refer to fixed points, this refers to fixed points in the parameter space of either $\{ \tilde{z}, \varphi\}$ in the phase space description or $\{s_x, s_y, s_z \}$ in the Bloch sphere description without including $n$.

In terms of the Bloch-sphere description, the fixed points are thus obtained by setting $\dot{s}_x = \dot{s}_y = \dot{s}_z = 0$ in Eqs.~\ref{eq:first zind}, \ref{eq:second zind} and \ref{eq:third zind}. Looking at the equation for $\dot{s}_z = 0$, one sees that the fixed point should satisfy either $s_y = 0$ or $s_z = J/2\delta J$. The latter solution implies $s_z > 1/2$ in the \pt-symmetric region, $J ^ 2 +\Delta^2 > \delta J^2$, which is not compatible with the condition $|\mathbf{s}|^2 = 1/4$ and thus not a valid solution. Therefore, the only fixed points in the \pt-symmetric region satisfy $s_y = 0$. Solving the other equations $\dot{s}_x = \dot{s}_y = 0$, we obtain two fixed points:
\begin{eqnarray}
\label{eq:zero_sy bloch zind}
\mqty(s_x\\ s_y\\ s_z )= 
\frac 1{2 (J^2 + \Delta^2)} 
\mqty( -\delta J\, \Delta \pm\Omega\,J \\ 
0 \\
\delta J\, J \pm \Omega\Delta ),
\end{eqnarray}
where $\Omega \equiv \sqrt{J^2 +\Delta ^2 -\delta J^2} > 0$ in the \pt-symmetric regime. 

Equivalently, the fixed points can be expressed in terms of the phase-space variables by setting $\dot{\tilde{z}} = \dot{\varphi} = 0$.
We find two fixed points,
\begin{eqnarray}
\label{eq:fixed zind 1}
\mqty( \tilde z\\ \varphi)= \qty{\mqty(\frac{\delta J\, J+\Omega\Delta}{J^2 + \Delta ^2} \\ 0 \text{ or }\pi),
\mqty( \frac{\delta J\, J-\Omega\Delta}{J^2 + \Delta ^2} \\ \pi) },
\end{eqnarray}
In the first fixed point, we should choose $\varphi = 0$ when $\delta J \!<\! J$ and $\varphi = \pi$ when $\delta J \!>\!J$.
This discontinuous change of $\varphi$ may look odd, but this is due to the fact that when $J = \delta J$, the first fixed point becomes $\tilde{z} = 1$, which corresponds to the north pole in the Bloch sphere description. As the fixed point crosses the north pole, the phase angle $\varphi$ changes discontinuously from $0$ to $\pi$. Note that there is no discontinuity in the Bloch sphere description. 

\begin{figure*}
	\qquad\quad 
	\includegraphics[width=.25\textwidth]{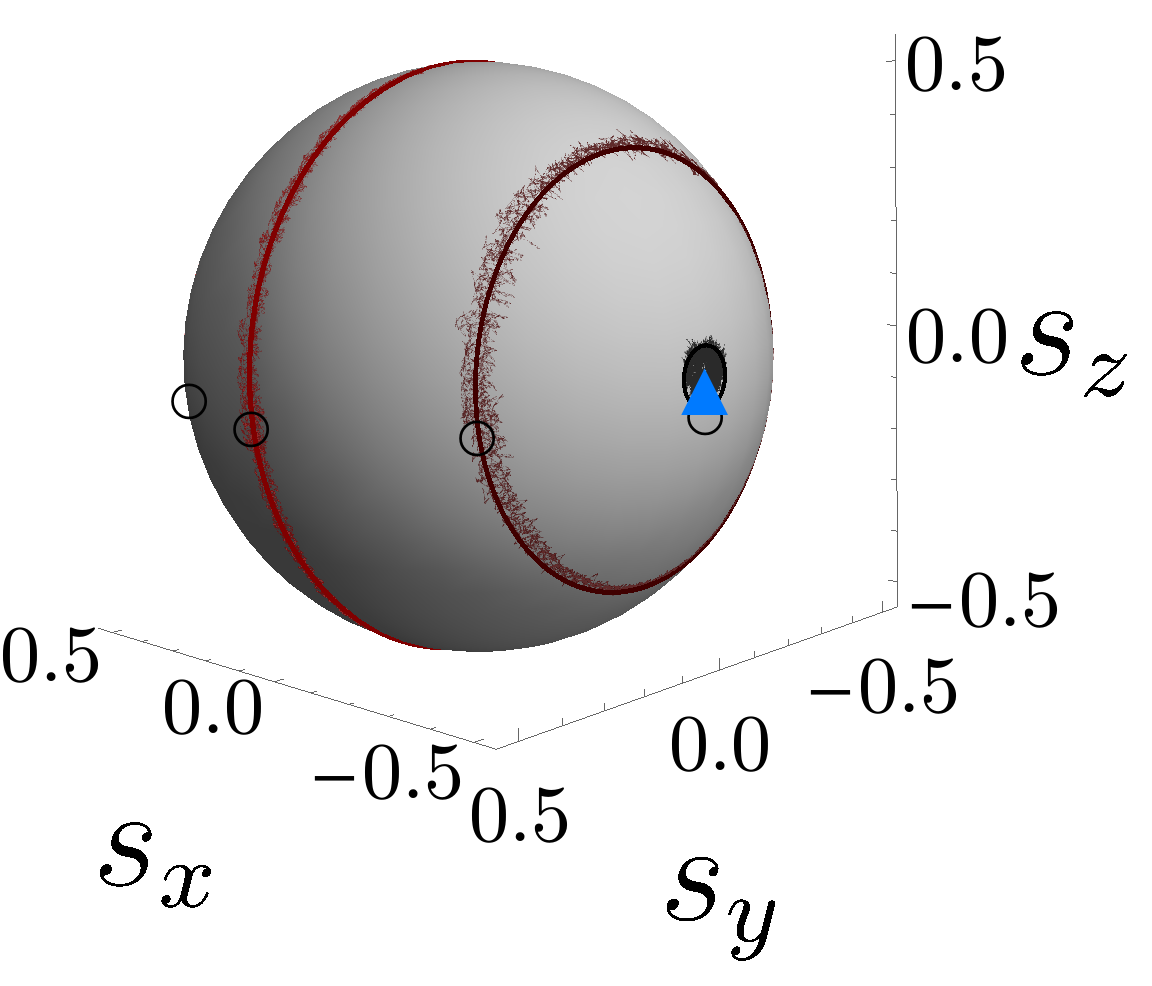}\qquad%
	\includegraphics[width=.25\textwidth]{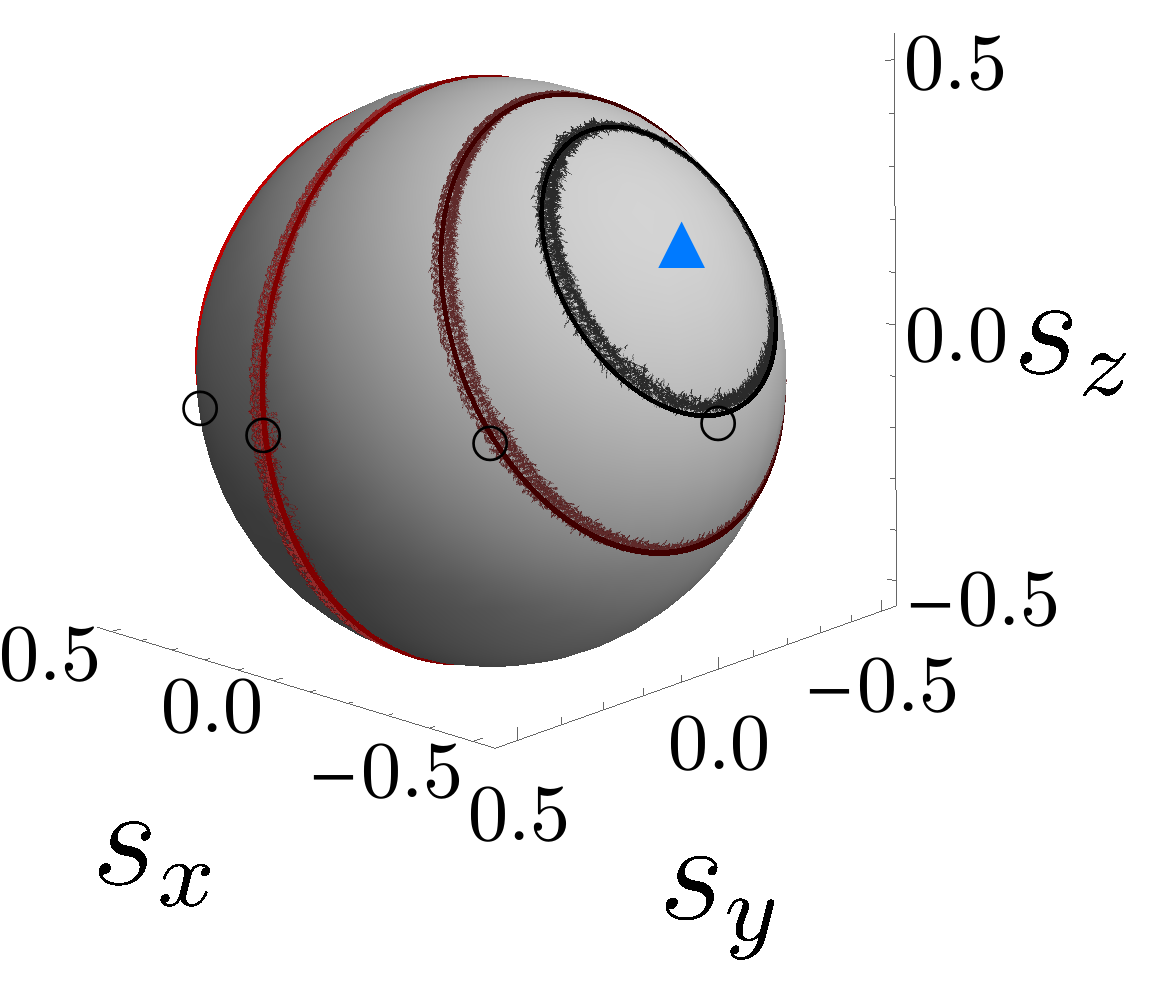}\qquad\,%
	\includegraphics[width=.25\textwidth]{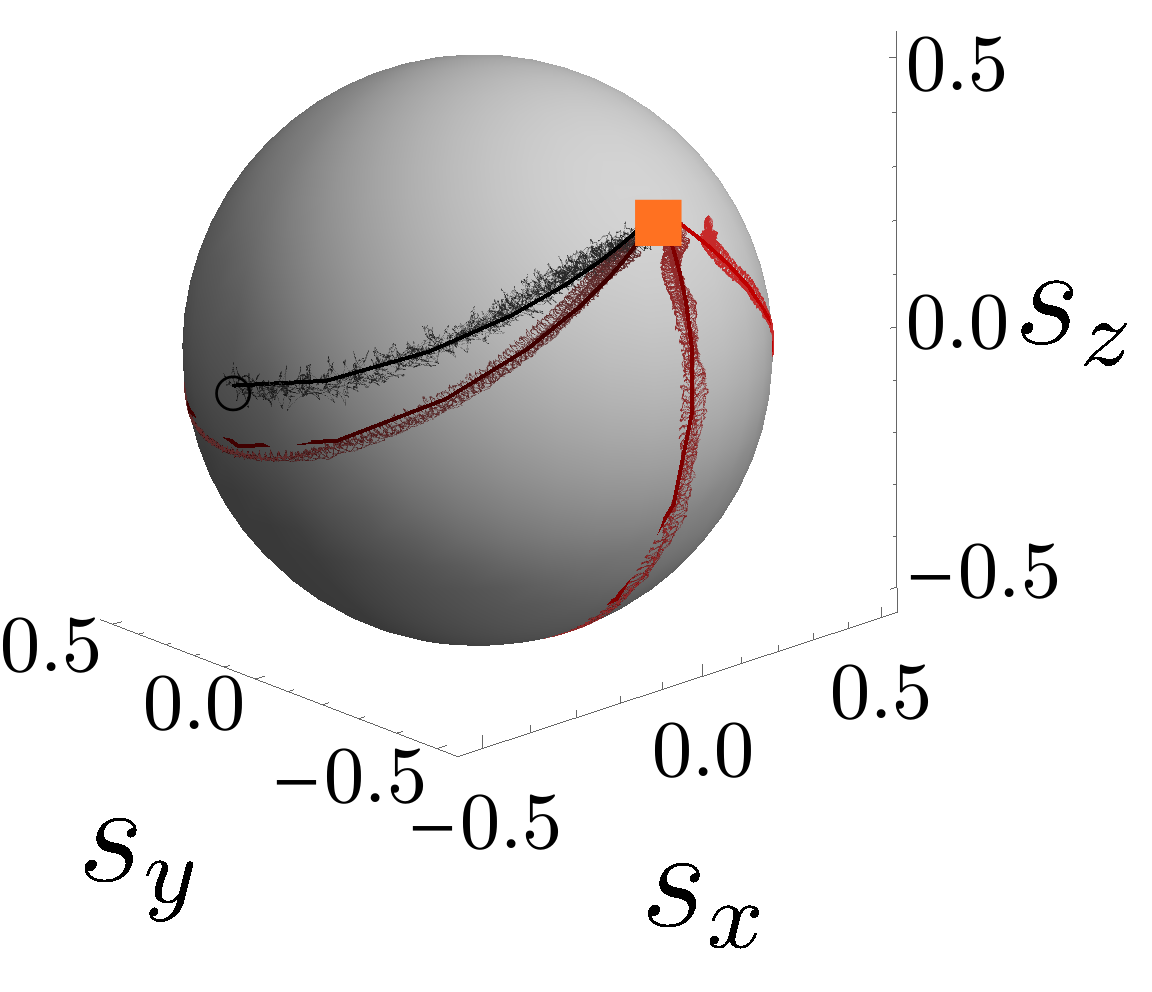}\\
	\includegraphics[width=.95\textwidth]{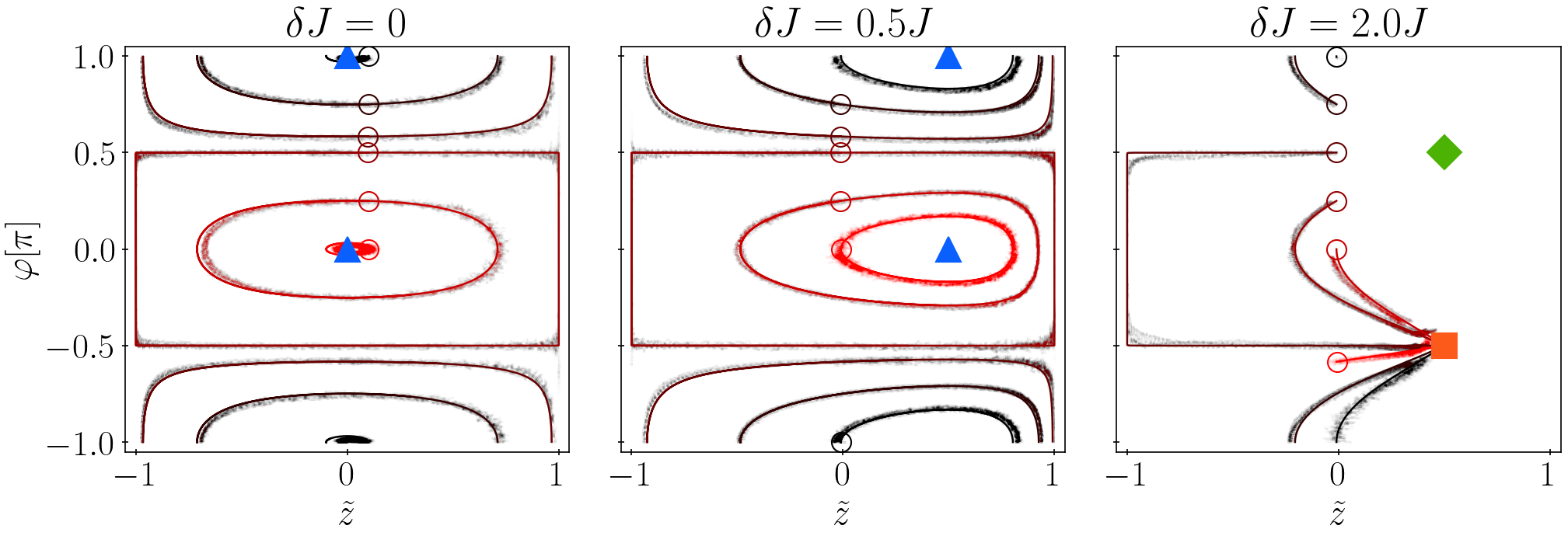}\qquad \qquad
 \caption{Numerically and experimentally obtained dynamics of the linear Hatano-Nelson dimer model, with $\Delta = 0 $, and three distinct values of $\delta J$ in both the Bloch-sphere representation (top panels) and the phase space representation (bottom panels). Different colors of points represent the different initial conditions, which are highlighted by open circles. Numerical results are in solid lines, whereas experimental data are plotted with (slightly transparent) points. In the \pt-symmetric regime the fixed points in Eqs.~\ref{eq:zero_sy bloch zind} and \ref{eq:fixed zind 1} are centers (blue triangles), while as soon as the \pt-symmetry is broken (panel on the right), one of the points in Eq.~\ref{eq:finite_sy bloch zind}, or equivalently in Eq.~\ref{eq:finite_sy zphin zind}, behaves as a source (unstable point, green diamond) and the other as a drain (stable point, orange square).}
 \label{fig:exp zind}
\end{figure*}

The dynamics around a fixed point can be understood by looking at the eigenvalues of the Jacobian matrix, $\mathcal J$, of the fixed point~\cite{hilborn, cencini,chong,zibold2010Classical}, which in the phase space $\{ \tilde{z}, \varphi \}$, is
\begin{eqnarray}
 \mathcal J =
 \begin{pmatrix}
   \partial_{\tilde z} \dot{\tilde z} & \partial_{\varphi} \dot{\tilde z}\\[1ex]
   \partial_{\tilde z} \dot{\varphi} & \partial_{\varphi} \dot{\varphi}
 \end{pmatrix}. \label{eq:jacobian}
 \end{eqnarray}
 If the real part of the two eigenvalues are both zero, the fixed point acts as a {\it center} of motion around which the system oscillates. If the real part of the eigenvalues are both positive, the fixed point is called an {\it unstable} fixed point, and the dynamics flows away from the point. If the real part of the eigenvalues are both negative, the fixed point is called a {\it stable} fixed point, and the dynamics sinks into the point. If one of the eigenvalues has a positive real part and the other eigenvalue has a negative real part, the fixed point behaves as a {\it saddle point} of dynamics.

The fixed points of Eq.~\ref{eq:fixed zind 1} behave as centers for the dynamics~\cite{graefe2010,graefe2012}.
This oscillatory behavior is analogous to the Rabi oscillation in coherently coupled two-level quantum systems. Such oscillatory dynamics is confirmed also experimentally, as seen in Fig.~\ref{fig:exp zind}. 
Depending on the value of $\delta J$, which is the strength of the non-Hermiticitiy, the value of $\tilde{z}$ at the fixed points varies, and it can even reach the maximum value of $\tilde{z} = 1$. This imbalance of population is a two-site version of the non-Hermitian skin effect known in the extended Hatano-Nelson model with edges, where all the eigenstates are known to be localized on one edge~\cite{hatano96,hatano97,hatano98}. 

These dynamical features are both numerically and experimentally confirmed, as described in Fig.~\ref{fig:exp zind}. The left panel describes the Hermitian limit ($\delta J = 0$) of the linear HN dimer model, whereas the central panel is in the \pt-symmetric regime and the right panel is the \pt-broken regime. Fixed points which serve as centers of dynamics are indicated by blue triangles, whereas the source and sink are indicated by a green diamond and an orange square, respectively. The roles of fixed points found in the dynamics obtained from numerical and experimental means are in accord with what we found above.

\begin{figure}[t!]
	\centering
	\includegraphics[width=
	\columnwidth]{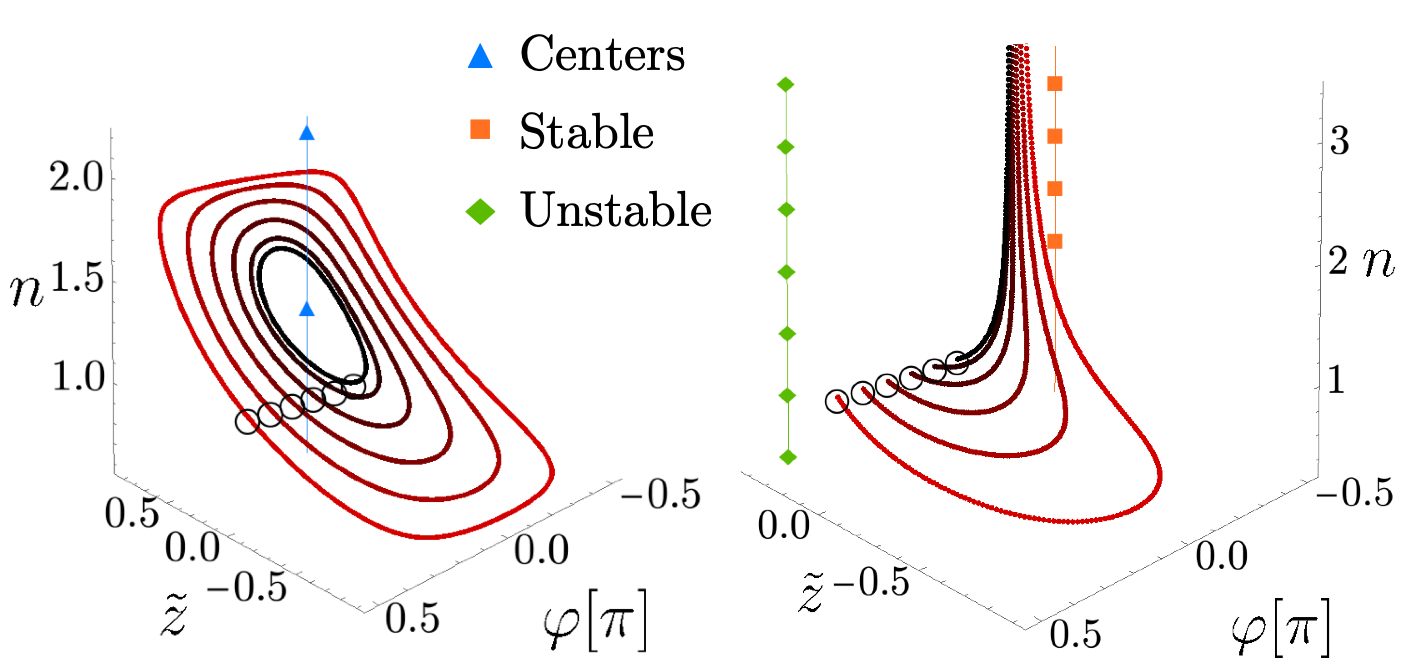}
	\caption{Numerically calculated phase-space dynamics for the linear HN dimer model in the space of $\{ n,\tilde{z},\varphi\}$. Different colors are for different initial conditions, highlighted with open circles. All the fixed points are represented as lines perpendicular to the $\tilde z-\varphi$ plane as their coordinates are independent of the population, $n$. Left panel: in the \pt-symmetric case ($\delta J =0.5 \, J$ and $\Delta = 0.01\, J$), the fixed points in Eq.~\ref{eq:fixed zind 1} (triangles) are centers. Right panel:  if the system is not \pt-symmetric ($\delta J = 2 \, J$ and $\Delta = 0.01\, J$), only points in Eq.~\ref{eq:finite_sy zphin zind} are well defined, and they appear in pairs of stable (orange squares) and unstable (green diamonds) points.}
	\label{fig:trajectories zind}
\end{figure}

If we insert the fixed points into the equation for $n$, we see $\dot{n} = 0$, indicating that the total population does not change in time on the fixed points. This behavior is expected because the eigenvalues of the Hamiltonian in the \pt-symmetric region are both real. However, if we look at the dynamics around the fixed points, the total population $n$ changes in time. In the left panel of Fig.~\ref{fig:trajectories zind}, we plot the numerically-calculated dynamics within a region of the 3D phase-space of $\{ n, \tilde{z}, \varphi \}$. In the 3D phase-space, the fixed point in the phase space $\{\tilde z, \varphi\}$ corresponds to a line perpendicular to the $\tilde{z}-\varphi$ plane. The variation of the population during dynamical evolution even in the \pt-symmetric region reflects that the eigenstates are not orthogonal, and their overlap can lead to a change of population during dynamics.

\subsubsection{\pt-broken Regime}

Approaching the \pt-breaking transition from the \pt-symmetric region corresponds to taking $\Omega \to 0$, and the fixed points of the \pt-symmetric region in Eq.(\ref{eq:zero_sy bloch zind}) merge in this limit, giving rise to an exceptional point~\cite{graefe2010,graefe2012}. 

Beyond the \pt-breaking transition, $J^2 +\Delta ^2 < \delta J^2$, we enter the \pt-broken regime where the eigenvalues of the Hamiltonian are no longer real.
The fixed point solution with $s_y = 0$ from the \pt-symmetric region is no longer a valid solution in the \pt-broken regime. Instead, there are two fixed points at
\begin{align}
\label{eq:finite_sy bloch zind}
\mqty(s_x\\ s_y\\ s_z )= 
 \frac 1{2\,\delta J} \mqty(-\Delta \\ \pm i\,\Omega \\ J)
.
\end{align}
Equivalently, in the phase-space description, the fixed points are:
\begin{eqnarray}
\label{eq:finite_sy zphin zind}
\mqty( \tilde z\\ \varphi)=\mqty( 
J/\delta J\\
\pm\arccos(- \Delta / \sqrt{\delta J^2 -J^2}) ).
\end{eqnarray}
The fixed point at $\varphi < 0$ is a stable fixed point acting as a sink of dynamics, whereas the fixed point at $\varphi > 0$ is an unstable fixed point acting as a source~\cite{graefe2010,graefe2012}.
In the linear HN dimer, fixed points are merely the eigenstates of the 2-by-2 Hamiltonian; the eigenstate corresponding to the stable fixed point has the eigenvalue whose imaginary part is larger than the other eigenstate. We can physically understand the source/sink nature of these fixed point by noting that, starting from a state that is a superposition of the two eigenstates, the weight of the eigenstate corresponding to the sink will grow because of the larger imaginary part. We note that the total population $n$ at the fixed points is no longer a constant, but rather grows in time. In this sense, these fixed points are not fixed points in the full dynamics taking $n$ into account. As we noted earlier in this paper, we keep this terminology that the fixed points are points where variables other than $n$ are kept constant.

\section{Instantaneous HN dimer model}
\label{sec:HN_inst}

Now we turn to models with nonlinearities. Unlike the linear HN dimer model where the fixed points correspond to the eigenstates of the 2-by-2 Hamiltonian, obtaining the fixed points of the dynamics of models with nonlinearity is generally nontrivial. Although the model we experimentally implement is the nonlinear HN model, this model does not allow for a simple description in terms of fixed points. However, as we shall explain, we can obtain the fixed points of the instantaneous HN model, which also contains nonlinearity and thus provides a good approximation to understand the experimentally realized nonlinear HN model. In this section, we give a theoretical description of the dynamics of the instantaneous HN dimer model, which we will use in later Sections to understand the dynamics of the nonlinear HN dimer model.

We rewrite here the equation of motion for the instantaneous HN dimer model Eq.(\ref{eq:Schroedinger inst}):
\begin{eqnarray}
 i\dot{\vb* \alpha} = 
\mqty( \omega -  \Delta & -J-\bar{g}\tilde{z} \\ -J+ \bar{g}\tilde{z} & \omega+\Delta) \vb* \alpha,
\end{eqnarray}
where we used the notation $\tilde{z} = z/n$ we introduced in the previous section.
The equations for the phase-space variables $\{ n,\tilde{z},\varphi\}$ are
\begin{align}
\dot n &= -2 \,\bar g n\, \tilde z \sqrt{1-\tilde z^2} \sin \varphi \label{eq:n dot2}\\
\dot{\tilde z} &= 2 (\bar g \,\tilde z^2 - J) \sqrt{1-\tilde z^2} \,\sin \varphi \label{eq:z dot2}\\
 \dot \varphi &= -2 \Delta +2 \frac{J-\bar g}{\sqrt{1-\tilde z^2}} \;\tilde z\cos\varphi, \label{eq:phi dot2}
\end{align}
while the equations for the Bloch sphere variables are:
\begin{align}
 \dot n &=  -8 \,\bar g n\,s_y s_z \label{eq:fourth2}\\
 \dot s_x  &= 2 \Delta s_y + 8 \,\bar g\, s_x s_y s_z \label{eq:first2}\\
 \dot s_y &= -2\Delta s_x  + 2 (J-\bar g) s_z + 8 \,\bar g \, s_y^2 s_z \label{eq:second2}\\
 \dot s_z &= -2J s_y + 8 \,\bar g \, s_y s_z^2. \label{eq:third2}
\end{align}
Just like the linear HN dimer model, time derivatives of the variables $\tilde{z}$, $\varphi$, $s_x$, $s_y$, and $s_z$ have no $n$ dependence, which implies that we can consider the dynamics in the phase space $\{ \tilde{z}, \varphi \}$ and on the Bloch sphere $\{ s_x, s_y, s_z\}$ independent from how $n$ depends on time. Nonetheless, the instantaneous HN dimer model is not a linear model; indeed the dynamics has $\tilde{z}$ dependence in the 2-by-2 nonlinear Hamiltonian of Eq.(\ref{eq:Schroedinger inst}). The situation should be contrasted with what happens for the nonlinear HN dimer model, whose equations of motions are Eqs.~(\ref{eq:n dot}-\ref{eq:third}) in the next section. The nonlinear HN dimer model has an explicit $n$ dependence in the equations of motion for $\tilde{z}$, $\varphi$, $s_x$, $s_y$, and $s_z$, and a closed description in terms of either $\{ \tilde{z}, \varphi \}$ or $\{ s_x, s_y, s_z\}$ is not allowed, making the analysis more difficult. The dynamics of the instantaneous HN dimer model we explore in this section serves as the basis to understand the dynamics of the nonlinear HN dimer model in the next section.

\subsection{Dynamics of the instantaneous HN dimer model}

As before, we first look for the fixed points of motion. Fixed points, as before, are points determined by $\dot{\tilde{z}} = \dot{\varphi} = 0$ for the phase-space dynamics and $\dot{s}_x = \dot{s}_y = \dot{s}_z = 0$ for the Bloch sphere dynamics, without imposing $\dot{n} = 0$. The fixed points we discuss are, therefore, fixed points in the restricted space in which we do not look at the time evolution of $n$.

From the condition $\dot{s}_z = 0$, we obtain $s_y = 0$ or $s_z^2 = J/(4\bar{g})$. Let us first examine the case $s_y = 0$. From the other two equations, $\dot{s}_x = \dot{s}_y = 0$, we obtain the following two fixed point solutions
\begin{align}
	\label{eq:aux zero_sy bloch}
	\mqty(s_x\\ s_y\\ s_z )= \pm \frac 1{2 \sqrt{(J-\bar g)^2 +\Delta ^2}}
	\mqty(J-\bar g \\ 
	0 \\
	\Delta).
\end{align}
A striking feature of the instantaneous HN dimer model is that these fixed points are always valid fixed points irrespective of the values of $J$, $\delta J$, and $\Delta$, which is to be contrasted with the linear HN dimer model where the fixed point $s_y = 0$ was valid only in the \pt-symmetric region.
These fixed points in the phase-space description are
\begin{align}
    \mqty(\tilde{z}\\ \varphi )=
    \qty{
    \mqty(
    \frac{\Delta\,\mathrm{sign}(J-\bar{g})}{\sqrt{(J - \bar{g})^2+\Delta^2}}
    \\
    0),
    \mqty(
    -\frac{\Delta\,\mathrm{sign}(J-\bar{g})}{\sqrt{(J - \bar{g})^2+\Delta^2}}
    \\
    \pi)
    }
\end{align}
The sign of the fixed points of $\bar{z}$ depends on
$\mathrm{sign}(J-\bar{g})$; this is because when $J = \bar{g}$, the fixed point reaches the north pole in the Bloch sphere description and thus $\varphi$ changes between 0 and $\pi$ around $J = \bar{g}$. Note that there is no discontinuous change in the Bloch sphere representation.

Now we look for the fixed point $s_y \neq 0$. Combining $s_z^2 = J/(4\bar{g})$ with $\dot{s}_x = \dot{s}_y = 0$, we obtain the following additional four fixed points
\begin{align}
	\mqty(s_x\\ s_y\\ s_z )= \pm \frac{1}{2\sqrt{\bar{g}J}}
	\mqty(
    -\Delta\\ 
	0 \\
	J) \pm \frac{1}{2\sqrt{\bar{g}}}
    \mqty(
    0\\ 
	\sqrt{\bar{g} - g_\mathrm{T}} \\
	0),
\end{align}
where the signs of the first and the second terms can be chosen independently, and we defined
\begin{align}
g_\mathrm{T} \equiv \frac{J^2 + \Delta^2}{J}.
\end{align}
These additional fixed points with $s_y \neq 0$ are valid fixed points only when $\bar{g} - g_\mathrm{T} > 0$.
Thus, the instantaneous model behaves qualitatively differently depending on whether $\bar{g} < g_\mathrm{T}$ or $\bar{g} > g_\mathrm{T}$, where the former has only two fixed points but the latter has six fixed points. We will refer to the case $\bar{g} < g_\mathrm{T}$ as the {\it weak interaction regime} and $\bar{g} > g_\mathrm{T}$ as the {\it strong interaction regime}.

These fixed points with $s_y \neq 0$ in the phase-space description have the following expressions:
\begin{align}
    \mqty(\tilde{z}\\ \varphi )=
	\mqty(
    \pm\sqrt{J/\bar{g}}\\ 
	\pm \arccos\left( \frac{\mp \Delta}{\sqrt{J(\bar{g}-J)}} \right)
	),
\end{align}
where the sign of $\tilde{z}$ and the sign inside $\arccos$ should be chosen to be opposite, but the sign in front of $\arccos$ is independent of the other signs.

We now consider the weak interaction regime $\bar{g} < g_\mathrm{T}$ and the strong interaction regime $\bar{g} > g_\mathrm{T}$ in turn to inspect the nature of fixed points and dynamics around them.

\subsubsection{Weak interaction regime}

In the weak interaction regime $\bar{g} < g_\mathrm{T}$, the instantaneous HN dimer model has two fixed points given by Eq.~(\ref{eq:aux zero_sy bloch}). These fixed points satisfy $\dot{n} = 0$, indicating that the total number does not change in time. Thus, the weak interaction regime is a direct analog of the \pt-symmetric regime in the linear HN dimer model.

We find that the eigenvalues of the Jacobian of the two fixed points in the phase-space description are $\lambda_\pm = \pm 2\sqrt{J(\bar{g} - g_\mathrm{T})}$ for both fixed points. Since $\bar{g} < g_\mathrm{T}$ in the weak interaction regime, $\lambda_\pm$ are complex conjugate pairs, indicating that these fixed points serve as the center of oscillation in the phase space of $\{ \bar{z},\varphi\}$, analogous to the fixed points in the \pt-symmetric regime in the linear HN dimer model.
\begin{figure}[b!]
	\centering
	\includegraphics[width=\columnwidth]{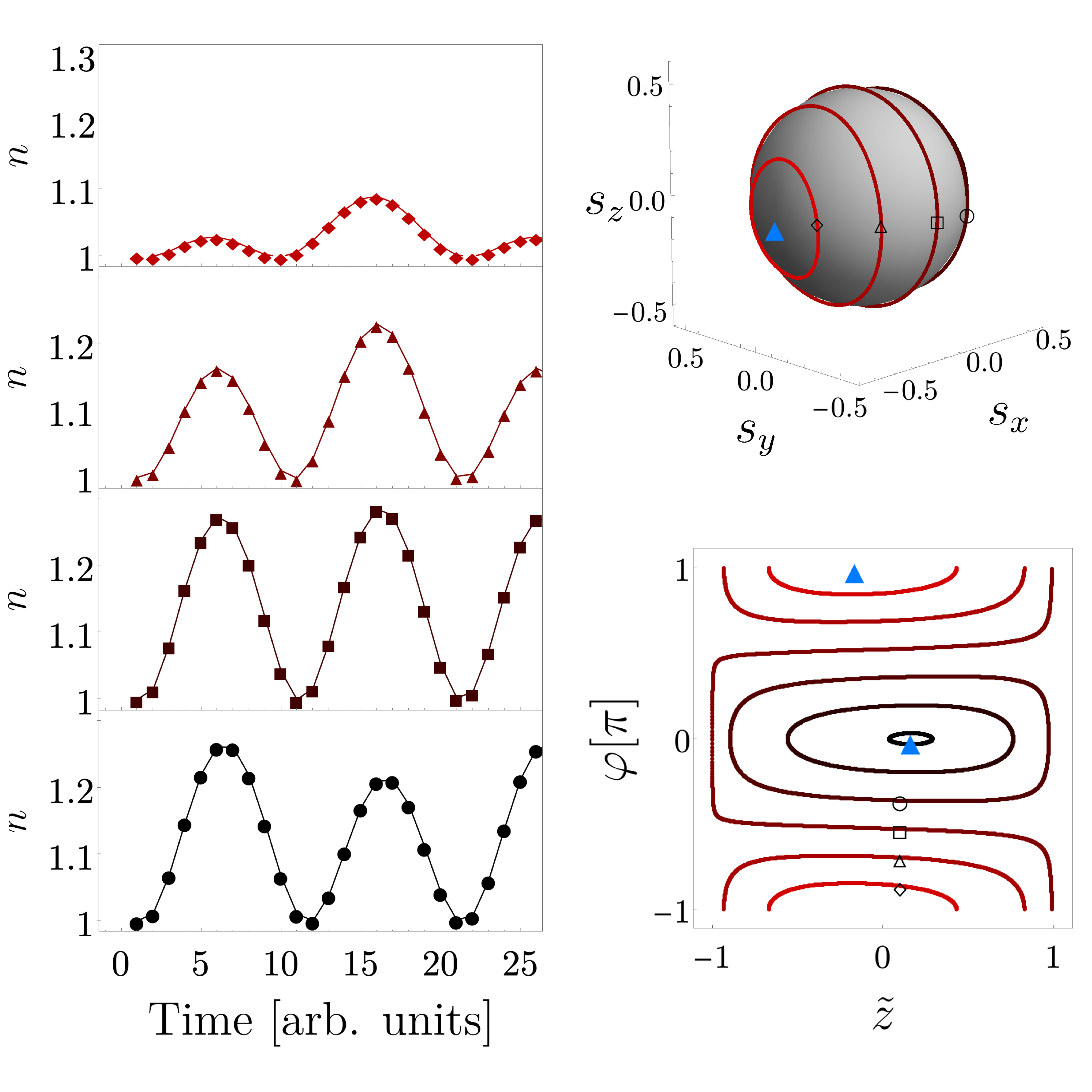}
	\caption{Numerically obtained dynamics for the instantaneous HN dimer model in the weak interaction regime $\bar g = 0.4 J <g_\mathrm T$ with $\Delta = 0.1 J$. The panels on the right show the dynamics on the Bloch sphere (top) and in the phase space (bottom) for a set of different initial conditions, as highlighted by the empty markers. The filled triangles in the bottom right plot represent the two fixed points which behave as centers. Rabi-like oscillations are clearly seen as is the shift of the fixed-point centers away from $\tilde{z}=0$, reflecting the interplay of the non-reciprocal couplings with the on-site energy difference. 
	The behavior of population in time for each of the initial conditions highlighted is plotted in the left panels.}
	\label{fig:aux1}
\end{figure}
In Fig.~\ref{fig:aux1}, we plot numerically obtained dynamics of the weak interaction regime. All orbits are closed as expected.

\subsubsection{Strong interaction regime}

The strong interaction regime $\bar{g} > g_\mathrm{T}$ of the instantaneous HN dimer model behaves differently from the \pt-broken regime of the linear HN dimer model. First of all, the two fixed points present in the weak interaction regime $\bar{g} < g_\mathrm{T}$ remain as fixed points in the strong interaction regime. The eigenvalues of the Jacobian of these two fixed points are $\lambda_\pm = \pm 2\sqrt{J(\bar{g} - g_\mathrm{T})}$, which are now both real and have opposite signs; this indicates that the fixed points now act as saddle points of dynamics in the $\{ \tilde{z}, \varphi \}$ phase space.

In addition to the two fixed points inherited from the weak interaction regime, there are four additional fixed points in the strong interaction regime. Calculating the eigenvalues of the Jacobian of these four fixed points, we can group them into two categories. 

Among the four additional fixed points, the eigenvalues of the Jacobian of the following two fixed points
\begin{align}
    \mqty(\tilde{z}\\ \varphi )=
	\mqty(
    \sqrt{J/\bar{g}}\\ 
	\arccos\left( \frac{-\Delta}{\sqrt{J(\bar{g}-J)}} \right)
	),\ 
    \mqty(
    -\sqrt{J/\bar{g}}\\ 
	-\arccos\left( \frac{\Delta}{\sqrt{J(\bar{g}-J)}} \right)
	),
\end{align}
are in both cases
\begin{align}
    \lambda = 2\sqrt{J(\bar{g} - g_\mathrm{T})} \text{ and } 4\sqrt{J(\bar{g} - g_\mathrm{T})},
\end{align}
which are both positive in the strong interaction regime $\bar{g} > g_\mathrm{T}$. Therefore, these two fixed points are unstable fixed points which act as sources of dynamics. 

\begin{figure*}[htb]
	\centering
  \includegraphics[width=.98\textwidth]{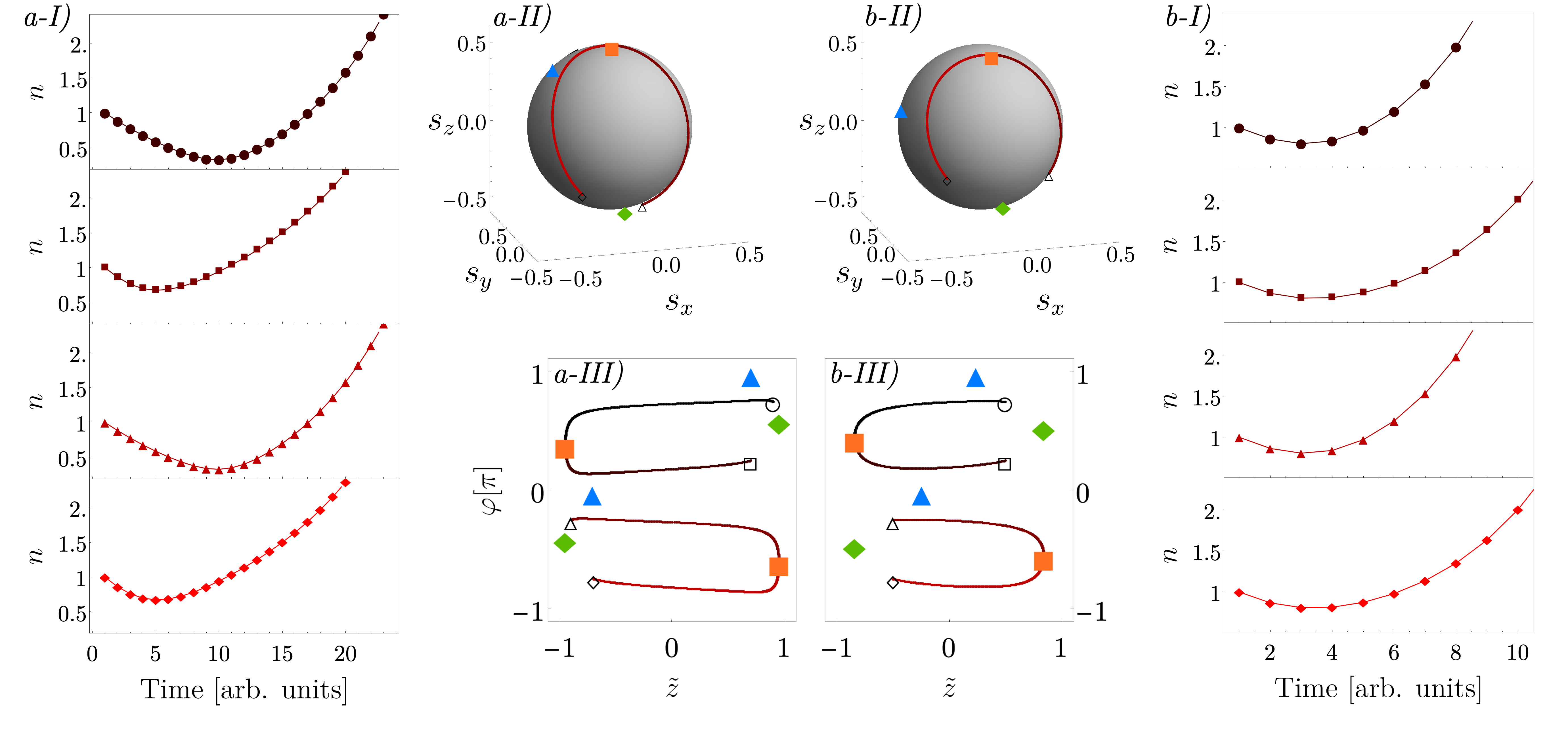}
	\caption{Numerically obtained dynamics of the instantaneous HN dimer model in the strong interaction regime with $\bar g = 1.1\,J > g_T$ (\textit{a}) and $\bar g = 1.4\, J$ (\textit{b}), both with $\Delta =0.1\,J$. In each case, we plot the behavior of the population over time (\textit{a-I} and \textit{b-I}), the trajectories on the normalized Bloch sphere (\textit{a-II} and \textit{b-II}) and the trajectories in the phase space of $\{ \tilde{z}, \varphi\}$ (\textit{a-III} and \textit{b-III}). Plots in the phase space and the Bloch sphere are for different initial conditions (marked with open markers); the evolution of the population over time for each initial condition is plotted in \textit{a-I} and \textit{b-I} using the corresponding marker. The saddle (blue triangles), stable (orange squares) and unstable (green diamonds) fixed points are highlighted. Note the trajectories converge towards the stable fixed points, after being slightly bent by the presence of saddle points. As expected, the population diverges quickly as a stable point is approached. 
 }
	\label{fig:aux2}
\end{figure*}

On the other hand, the eigenvalues of the Jacobian of the other two fixed points
\begin{align}
    \mqty(\tilde{z}\\ \varphi )=
	\mqty(
    \sqrt{J/\bar{g}}\\ 
	-\arccos\left( \frac{-\Delta}{\sqrt{J(\bar{g}-J)}} \right)
	),\ 
    \mqty(
    -\sqrt{J/\bar{g}}\\ 
	\arccos\left( \frac{\Delta}{\sqrt{J(\bar{g}-J)}} \right)
	),
\end{align}
are both
\begin{align}
    \lambda = -2\sqrt{J(\bar{g} - g_\mathrm{T})} \text{ and } -4\sqrt{J(\bar{g} - g_\mathrm{T})},
\end{align}
which are both negative, indicating that these two fixed points are stable fixed points that act as sinks of dynamics. 
This behavior is confirmed by the numerically obtained dynamics as plotted in Fig.~\ref{fig:aux2}.

Calculating the time dependence of the total population at these four fixed points, we find
\begin{align}
\dot{n} = \pm 2n\sqrt{J(\bar{g} - g_\mathrm{T})},
\end{align}
where the positive (negative) sign corresponds to the two stable (unstable) fixed points.
This indicates that the total population of the stable fixed points exponentially increases, whereas that of the unstable fixed points exponentially decreases; such behavior is consistent with these stable points acting as sinks and sources of dynamics. 

For all four of these fixed points, the population imbalance is $|\bar{z}| = \sqrt{J/\bar{g}}$, which indicates that $|\bar{z}| \to 0$ as $\bar{g} \to \infty$. The non-Hermitian localization on one of the two sites becomes smaller and smaller as the interaction $\bar{g}$ becomes larger, which is in stark contrast to the self-trapping phenomenon known in the two-site nonlinear Josephson model where the localization becomes stronger as the interaction becomes larger.

\subsubsection{Transition between the weak and strong regimes}

We have just seen that, as the interaction $\bar{g}$ increases and crosses $g_\mathrm{T}$, the number of fixed points changes from two to six. We now examine this transition. 

We first note that such an increase of the number of fixed points beyond two is not possible with any linear dimer model, in which the fixed points are determined by (at most) two eigenstates of the two-by-two Hamiltonian. Therefore, the appearance of six fixed points in the strong regime is an intrinsically nonlinear phenomenon.

Approaching the transition point from the weak regime, each of the two fixed points split into three as one crosses the transition point $\bar{g} = g_\mathrm{T}$. During this process, a fixed point on the weak interaction side, which is a center of dynamics, turns into three fixed points which are a saddle point, a stable fixed point, and an unstable fixed point. This process is consistent with the {\it Poincar\'e-Hopf index theorem} as we shall explain. On a two-dimensional parameter space, such as the phase space $\{ \bar{z}, \varphi\}$ and the surface of the Bloch sphere, the tangent vectors of the dynamics define a vector field. Such a vector field can have singularities, corresponding to the fixed points. For each of these singularities, a topological index called the Poincar\'e index can be defined, which assigns the value of $-1$ for saddle points and $+1$ for centers, stable, and unstable fixed points. The index theorem states that the sum of the Poincar\'e indices on the two-dimensional parameter space should be equal to the Euler characteristics of the parameter space~\cite{Arnold, graefe2010}. The Euler characteristics of our parameter space, which is a two-dimensional sphere as evident from the Bloch sphere description, is $+2$. In the weak interaction regime, we have two centers as singularities, and thus the sum of the Poincar\'e indices is $+2$, which is equal to the Euler characteristics as expected. As one crosses the transition point, a center, which has the Poincar\'e index of $+1$, turns into a saddle point, a stable point, and an unstable point, whose Poincar\'e indices are $-1$, $+1$, and $+1$, respectively, conserving the sum of the Poincar\'e indices. Thus the strong interaction regime also satisfies the index theorem.

\section{Nonlinear HN dimer model}
\label{sec:experimental model}

With the understanding of the instantaneous HN dimer model, we can now understand the dynamics of the nonlinear HN dimer model, which is the model experimentally implemented.

We rewrite the equations of motion for the nonlinear HN dimer model Eq.~(\ref{eq:Schroedinger}),
\begin{eqnarray}
\label{eq:Schroedinger exp}
i\frac{\partial }{\partial t} \left(\begin{array}{c} \alpha_1 \\ \alpha_2\end{array} \right) = 
\mqty( \omega -  \Delta & -J - g z  \\ -J+ g z & \omega+\Delta) \left(\begin{array}{c} \alpha_1 \\ \alpha_2 \end{array}  \right).
\end{eqnarray}
The dynamical equations in the phase-space are
\begin{align}
\dot n &= -2 \,g n^2 \,\tilde z\, \sqrt{1-\tilde z^2} \sin \varphi, \label{eq:n dot} \\
\dot {\tilde z} &= 2\, (gn \,\tilde z^2 - J)\, \sqrt{1-\tilde z^2} \,\sin \varphi, \label{eq:z dot}\\
 \dot \varphi &= -2\, \Delta +2 \,\frac{J-gn}{\sqrt{1-\tilde z^2}} \;\tilde z\cos\varphi. \label{eq:phi dot}
\end{align}
and in terms of the Bloch sphere variables, the dynamics obey
\begin{align}
 \dot n \;&=  -8 gn^2\,s_y s_z, \label{eq:fourth}\\
 \dot s_x  &= 2 \Delta s_y + 8 gn\, s_x s_y s_z, \label{eq:first}\\
 \dot s_y &= -2\Delta s_x  + 2 (J-gn) s_z + 8 g n \, s_y^2 s_z ,\label{eq:second}\\
 \dot s_z &= -2J s_y + 8 g n \, s_y s_z^2 . \label{eq:third}
\end{align}
We see that the time dependence of $\tilde{z}$, $\varphi$ and $s_x$, $s_y$, $s_z$ depend explicitly on $n$, implying that we can no longer use the two-dimensional phase space $\{\tilde{z}, \varphi\}$ and the surface of the Bloch sphere $\{ s_x, s_y, s_z\}$ as a parameter space within which the equations of motion are closed. Instead, we should consider the time dependence of $n$ together with the other variables to understand the dynamics. This explicit $n$ dependence in the equations of motion makes it difficult to analytically approach the dynamics of the nonlinear HN dimer model when the interaction is strong. However, as we shall see, the theory of the instantaneous HN dimer model can provide a good qualitative understanding of what happens in the nonlinear HN dimer model.

\subsection{Dynamics of the nonlinear HN dimer model}
We first look for fixed points of the dynamics. Since replacing $\bar{g}$ in the instantaneous HN dimer model by $gn$ recovers the equations of motion of the nonlinear HN dimer model, the two fixed points in the weak interaction regime and the six fixed points in the strong interaction regime of the instantaneous HN dimer model still satisfy $\dot{\tilde{z}} = \dot{\varphi} = 0$ and $\dot{s}_x = \dot{s}_y = \dot{s}_z = 0$. However, these points are not truly fixed anymore in the parameter space because $n$ can depend on time, and the time dependence of $n$ itself affects the position of the fixed points in the parameter space.

We first note that the two fixed points satisfying $s_y = 0$ are still fixed in the parameter space even in the nonlinear HN dimer model because they obey $\dot{n} = 0$, namely $n$ is time independent. The additional four fixed points are not fixed anymore in the parameter space because $n$ changes in time. The transition between the weak and strong interaction regimes is clear in the case of the instantaneous HN dimer model, given by $\bar{g} = g_\mathrm{T}$. On the other hand, in the case of the nonlinear HN dimer model, there appears also a distinctive intermediate interaction regime between the weak and the strong regimes as we shall explain now.
 
In the nonlinear HN dimer model, one measure of the interaction strength is $g n_0$, where $n_0$ is the total population at the initial time. Replacing $\bar{g}$ by $g n_0$, the two fixed points with $s_y = 0$ of the instantaneous HN dimer model are still the fixed points in the nonlinear HN dimer model. If $g n_0$ is small enough compared to $g_\mathrm{T}$, the dynamics we found for the weak interaction regime of the instantaneous HN dimer model applies also to the nonlinear HN dimer model. However, one should remember that, although the total population $n$ does not change in time exactly at the fixed points, the total population does change during the periodic Rabi oscillation around the fixed points. This implies that even though the system initially satisfies $g n_0 < g_\mathrm{T}$, the total population changes and at some time $t$ we may enter the regime with $gn(t) > g_\mathrm{T}$ at which the dynamics should be compared to the strong interaction regime of the instantaneous HN dimer model.

We refer to the regime in which $g n(t) < g_\mathrm{T}$ holds for any time $t \ge 0$ as the weak interaction regime of the nonlinear HN dimer model. In the weak interaction regime, all the dynamics are described by orbital motion around the two fixed points which serve as the centers of dynamics.

\begin{figure*}[ht!]
	\centering
	\includegraphics[width=.22\textwidth]{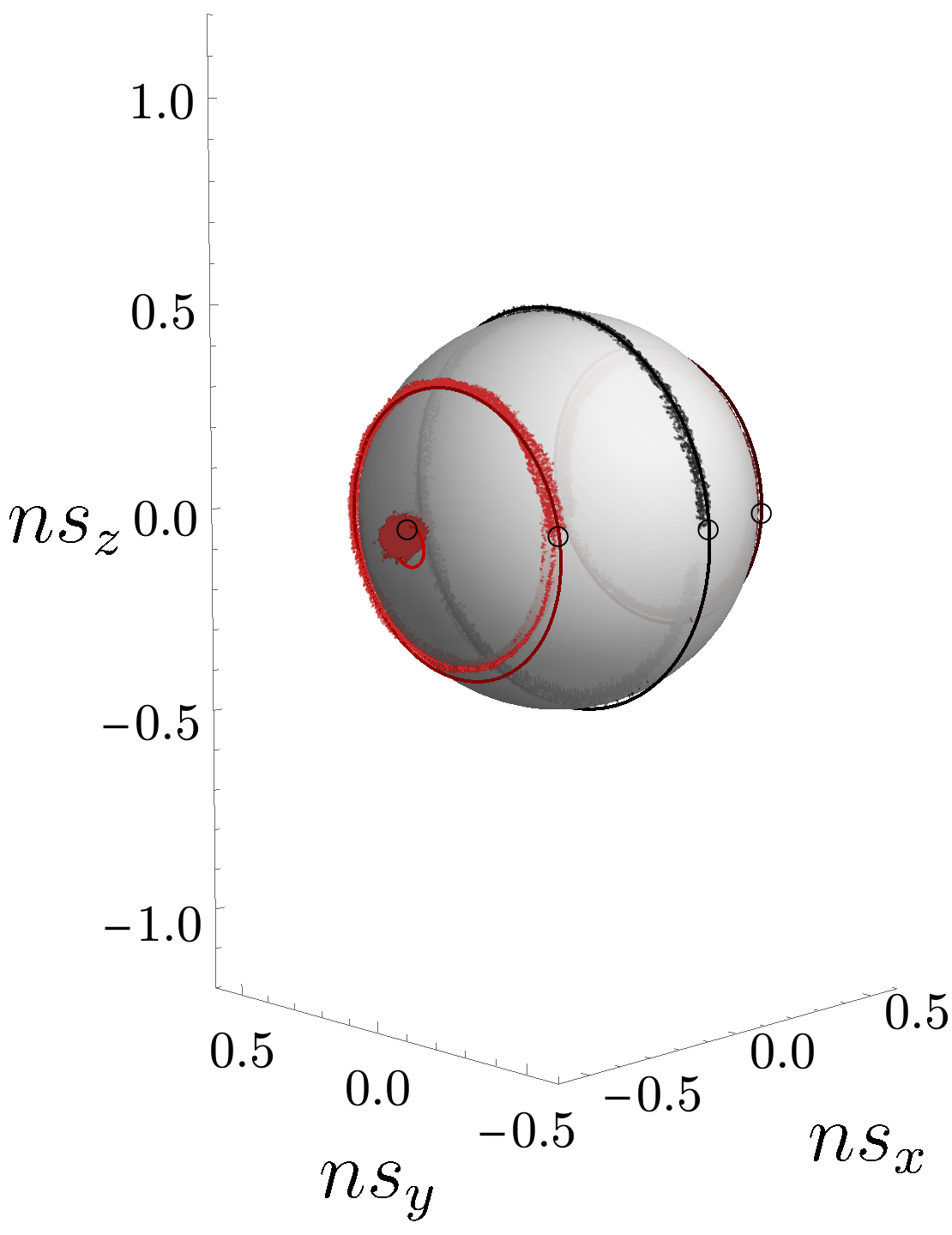}%
	\includegraphics[width=.22\textwidth]{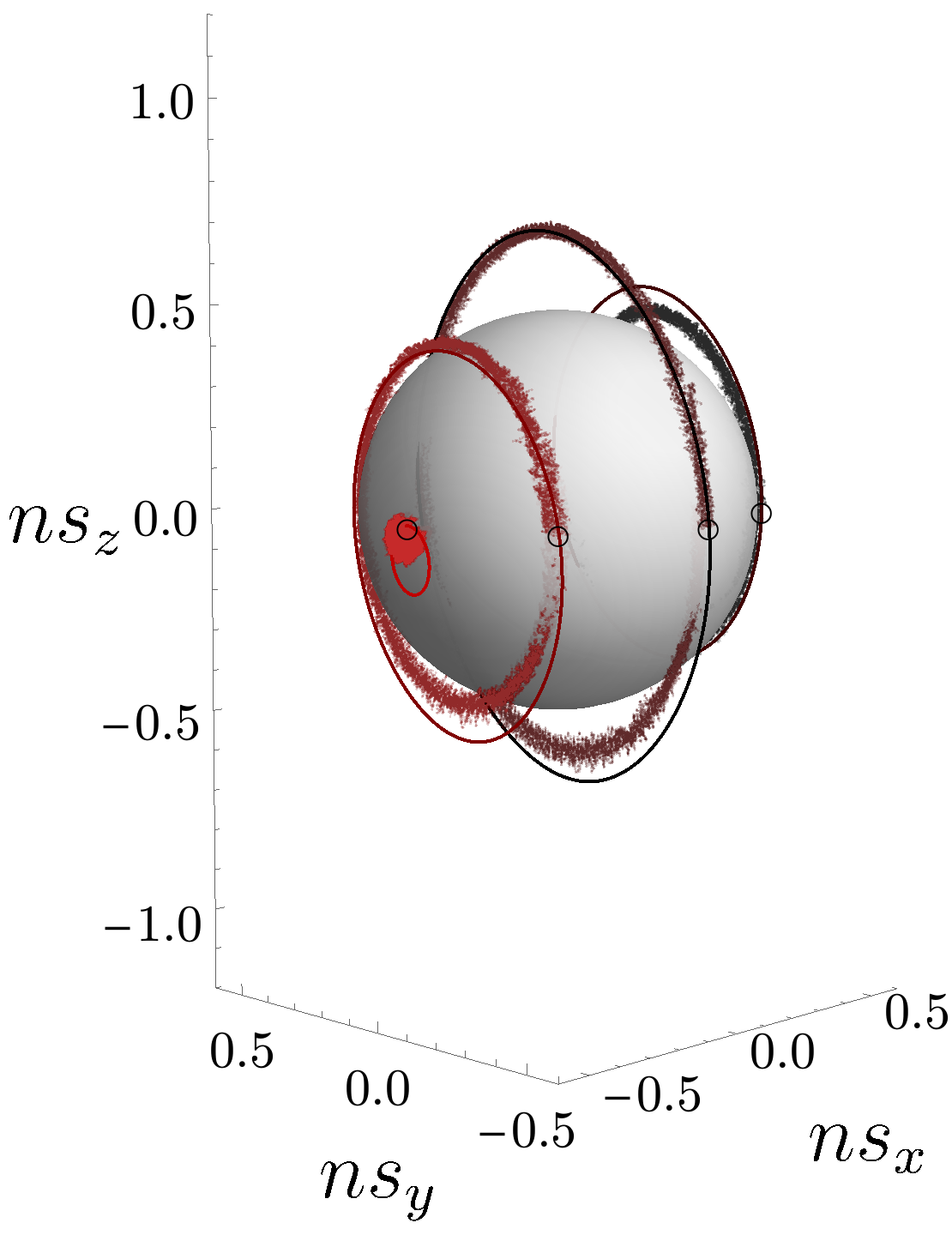}%
	\includegraphics[width=.22\textwidth]{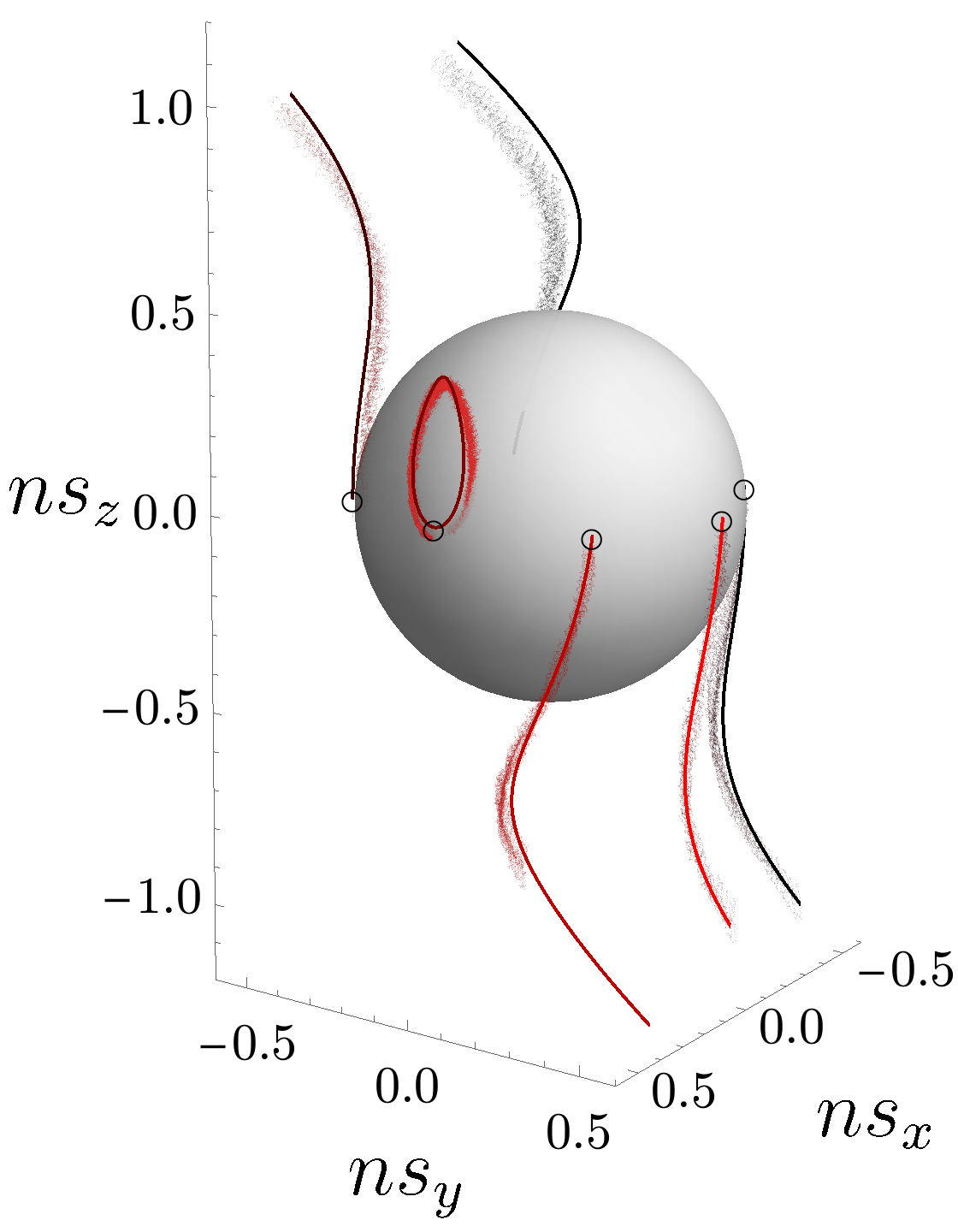}%
	\includegraphics[width=.22\textwidth]{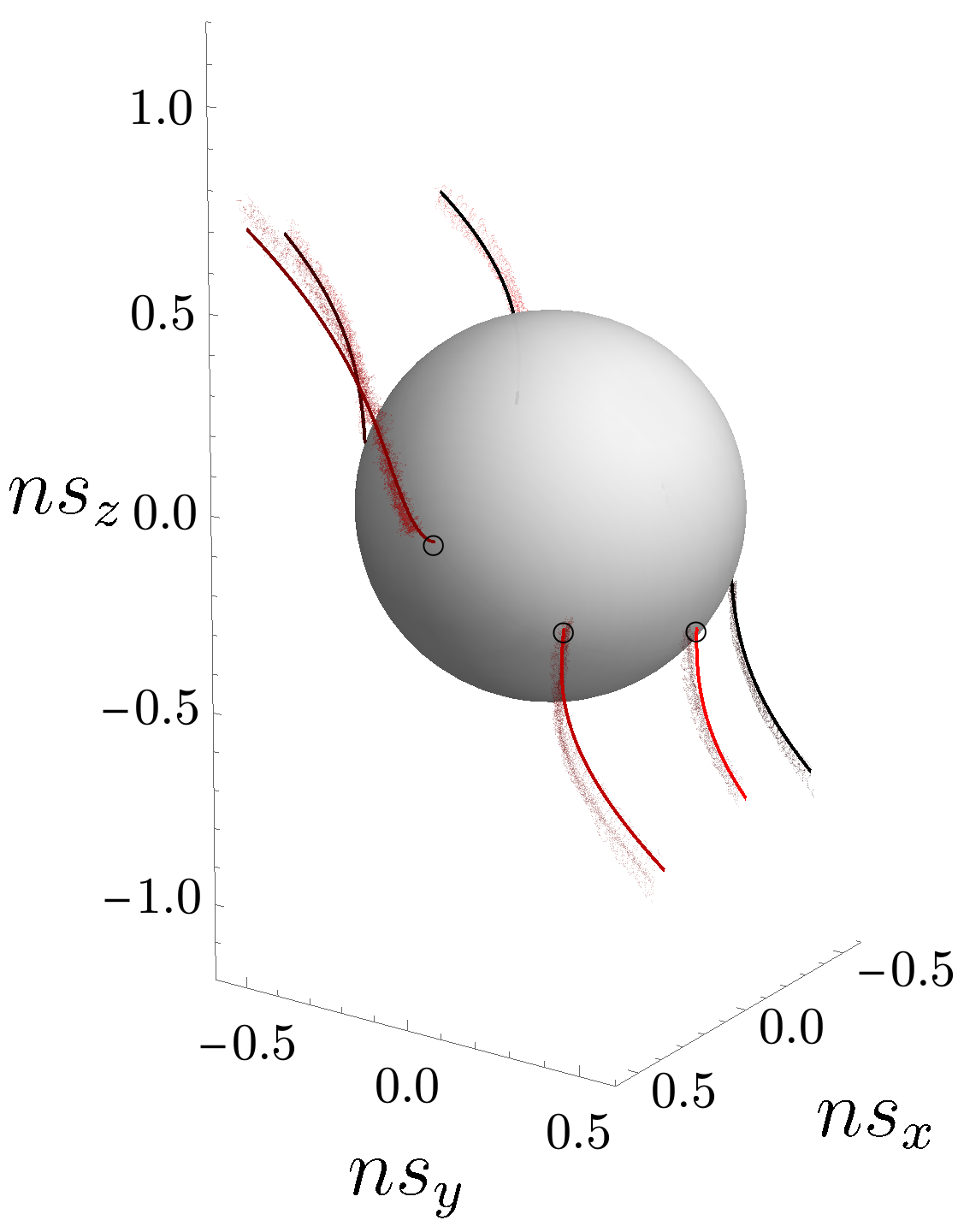}\\
	\includegraphics[width=.95\textwidth]{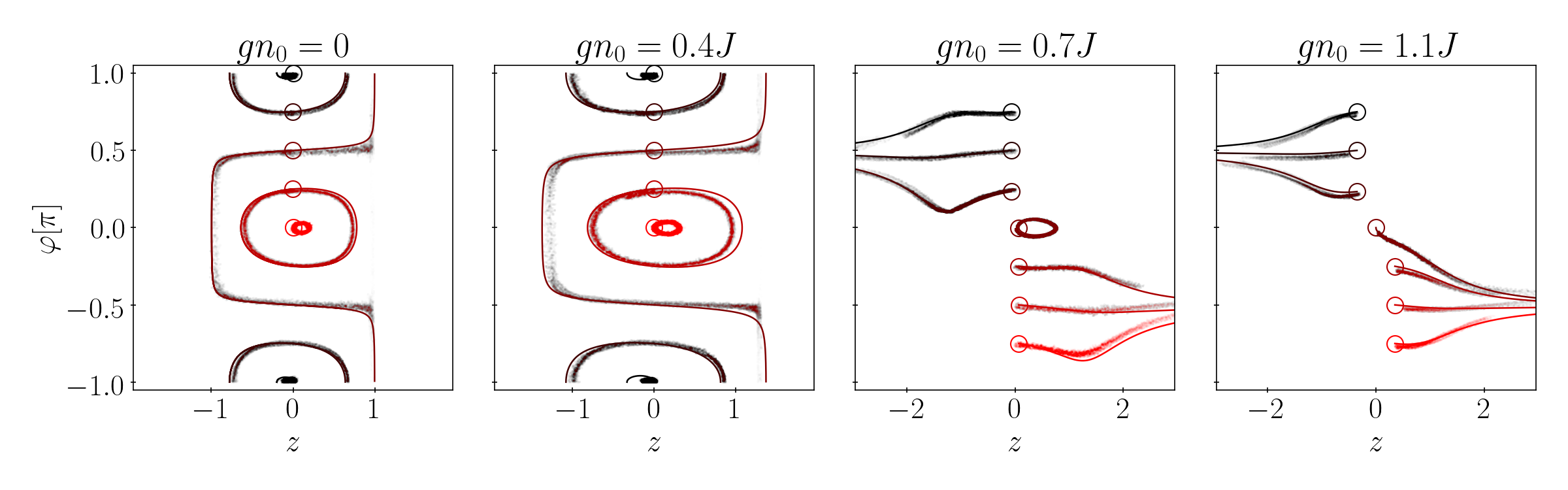} \qquad
	\caption{Comparison between the numerical simulation and the experimental results of dynamics of the nonlinear HN dimer model, for $\Delta = 0.1 J $, and four distinct values of $g n_0$ in both the {\it unnormalized} Bloch-sphere representation (top panels), in which we plot $ n \vb s = n (s_x, s_y, s_z)^\textrm T$, and the phase space representation (bottom panels). Numerical results are in solid lines, whereas experimental data are plotted with (slightly transparent) points, open circles represent the initial conditions. We note that the points can leave the surface of the Bloch sphere because the distance from the origin reflects the total population $n$. Three regimes are identified: for zero or small values of $g$ (two panels on the left), Rabi-like oscillations are observed just like in Fig.~\ref{fig:exp zind}. In the opposite limit of very strong interactions (right-most panel), the system tends towards an asymptotic value of $z$, either positive or negative. In the intermediate regime, $g n_0 = 0.7J$, the two behaviors coexist, depending on the initial conditions. }
	\label{fig:exp z dep}
\end{figure*}

We refer to the regime in which both $g n(t) < g_\mathrm{T}$ and $g n(t) > g_\mathrm{T}$ happen at some time $t \ge 0$ during the evolution as the {\it intermediate} interaction regime. In this intermediate interaction regime, which is a unique feature of the nonlinear HN dimer model, there simultaneously exist two types of orbits: one is a closed orbit similar to the weak interaction regime, and the other is a diverging orbit which is reminiscent of the strong interaction regime of the instantaneous HN dimer model.

We finally refer to the regime in which $g n(t) > g_\mathrm{T}$ holds for any time $t \ge 0$ as the strong interaction regime, where the dynamics diverges as in the strong interaction regime of the instantaneous HN dimer model. A crucial difference between the strong interaction regime of the instantaneous HN dimer model and the nonlinear dimer model is that, in the latter, the source and the sink of the dynamics are no longer fixed points in the space $\{\tilde{z}, \varphi\}$ or $\{ s_x, s_y, s_z\}$ because of the change of the total population $n$.

We now examine these three regimes experimentally, and we compare them to numerical simulations.

\subsection{Experimental results and Numerical simulations}

Figure~\ref{fig:exp z dep} shows the dynamics of the nonlinear HN dimer model, experimentally and numerically obtained, for different values of $gn_0$.
Firstly, panels with $g n_0 = 0$ and $g n_0 =0.4 \,J$ correspond to the weak interaction regime. We observe Rabi-like oscillations, similar to the ones found in the linear and instantaneous HN dimer models.
Secondly, the panel with $g n_0 = 1.1 J$ corresponds to the strong interaction regime, where the population of the trajectories diverges over time towards either positive or negative population imbalances, depending on the initial conditions.
Finally, in the intermediate interaction regime at $g n_0 = 0.7 J$ the two behaviors co-exist, \textit{i.e.}, stable orbits and unstable trajectories are possible, depending on the initial conditions used, i.e. the initial phase difference between the two sites. This co-existence of both regimes is the unique feature of the nonlinear HN dimer model.

\begin{figure*}[t]
	\centering
	\includegraphics[width=.3\textwidth]{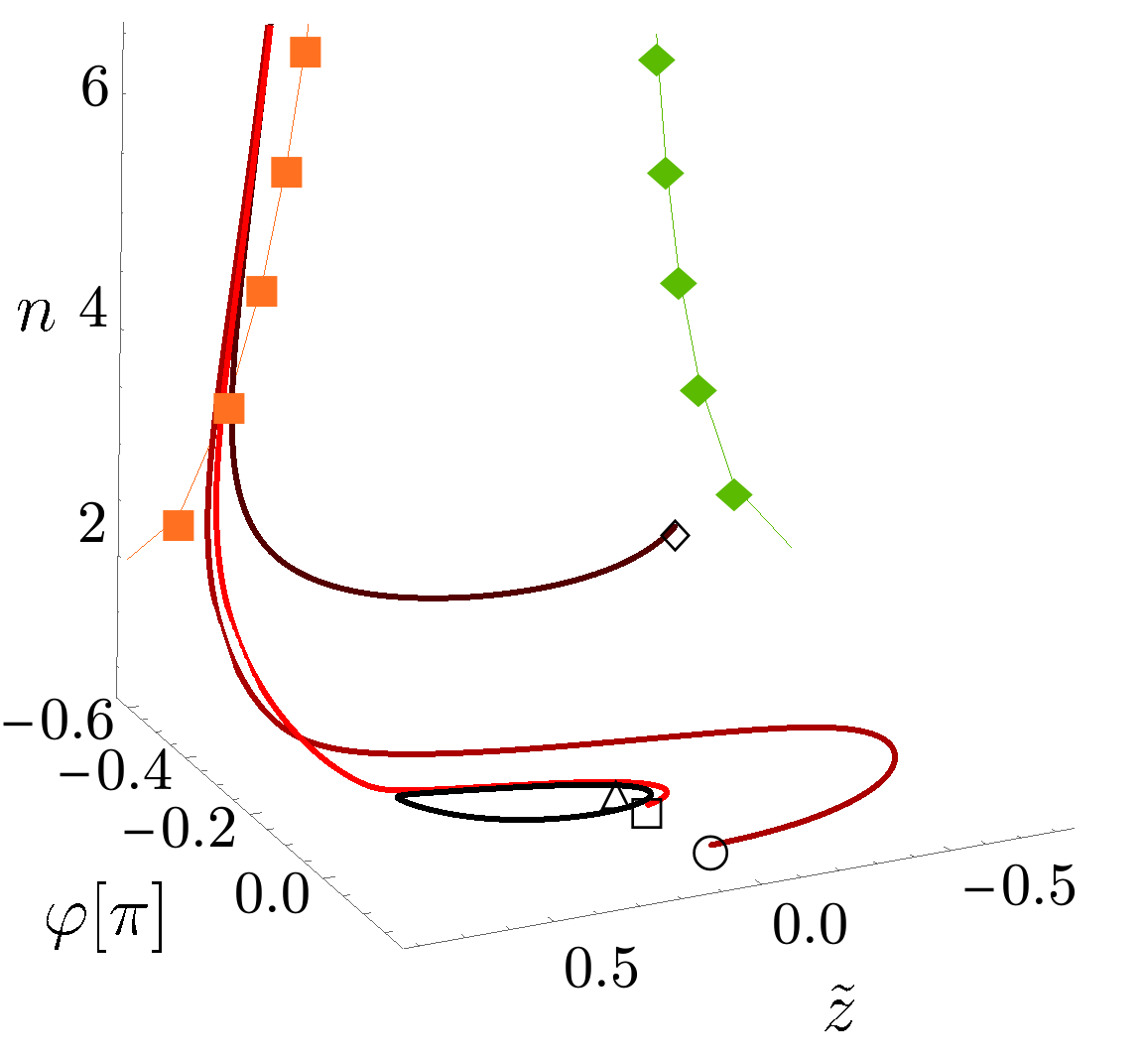}%
	\includegraphics[width=.3\textwidth]{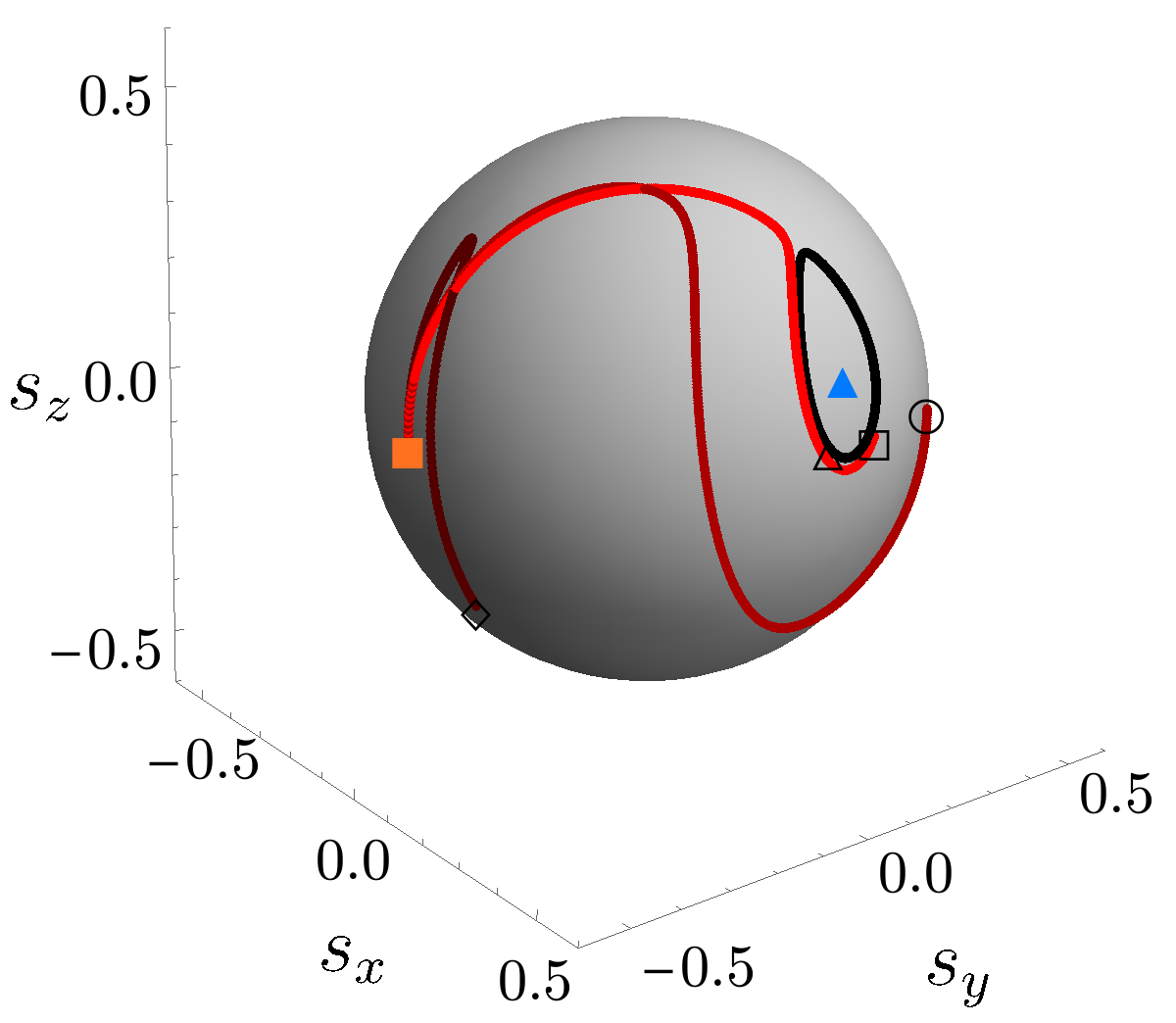} %
	\includegraphics[width=.39\textwidth]{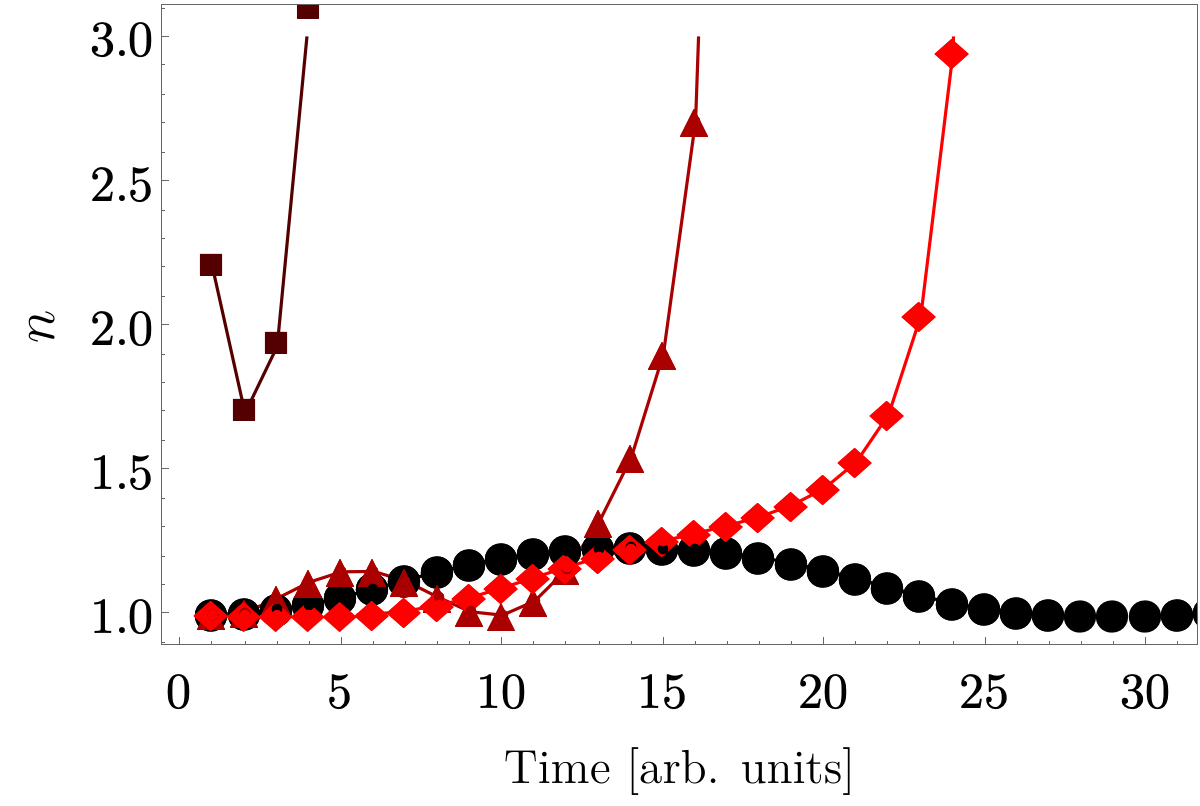} %
	\caption{Numerical dynamics of the population in the phase-space for the nonlinear HN dimer model with $g = 0.7 J$ and $\Delta = 0.1\,J$; initial conditions are chosen so to highlight the different behaviors described in the text. The empty markers in the phase space plot and Bloch sphere plot (left and central panel) highlight the initial conditions, while different colors are for trajectories starting from each of the initial conditions; the change of $n$ over time of each set of initial conditions (right panel, corresponding markers and colors) shows an evident decrease of population in the neighborhood of the unstable points (green diamonds). As the system evolves, it is then attracted towards the stable points (orange squares) where $n$ grows exponentially.}
		\label{fig:traj2}
\end{figure*}

We can further see the unique feature of the nonlinear HN dimer model if we examine the dynamics on the normalized Bloch sphere, as shown in the central panel of Fig.~\ref{fig:traj2}. 
We observe that the trajectories bend in arcs on the sphere, before eventually converging towards points on the equator. This behavior arises because instantaneously the trajectory is attracted towards the stable fixed points of the instantaneous HN dimer model for that particular value of $g n \rightarrow \bar{g}$. However, as $n$ keeps growing, the coordinates of the fixed points in the instantaneous HN dimer model also keep changing (see, \textit{e.g.}, Fig.~\ref{fig:aux2}), so that the trajectories appear to effectively ``chase down" the stable points by following these arcs. Indeed, in the limit that $n \rightarrow \infty$, the coordinates on the Bloch sphere of the stable fixed points in the instantaneous HN dimer model become:
\begin{eqnarray}
\begin{aligned}
\mqty(s_x \\  s_y \\  s_z) \rightarrow 
\mqty(0 \\    \pm {1/2} \\  0),
\end{aligned}
\end{eqnarray}
corresponding to the points of convergence on the equator of the normalized Bloch sphere in the experimental model, \textit{e.g.}, as can be seen in the central panel of Fig.~\ref{fig:traj2}. Physically, this corresponds to the modes changing from being localized primarily on one of the two sites at small $\bar{g}$ (and hence small $n$), to being an equal superposition of the two sites as $\bar{g} \rightarrow \infty$, as the non-reciprocal coupling term dominates over all other terms in the Hamiltonian.

Of particular note in Fig.~\ref{fig:traj2}, two trajectories starting close to each other (black points and red points in the top panels) actually behave very differently: one describes a closed loop around the center (full blue triangle) and its population remains finite, although oscillating, whereas the other about half-way through the loop, diverts and ends up in the basin of attraction of the stable points (full orange squares), where the population diverges exponentially. As in the strong interaction case, note that the trajectories are bent in arcs to ``follow" the always changing position of the fixed point in the corresponding instantaneous model.

\begin{figure*}
	\centering
	\includegraphics[height=.37\textwidth,clip]{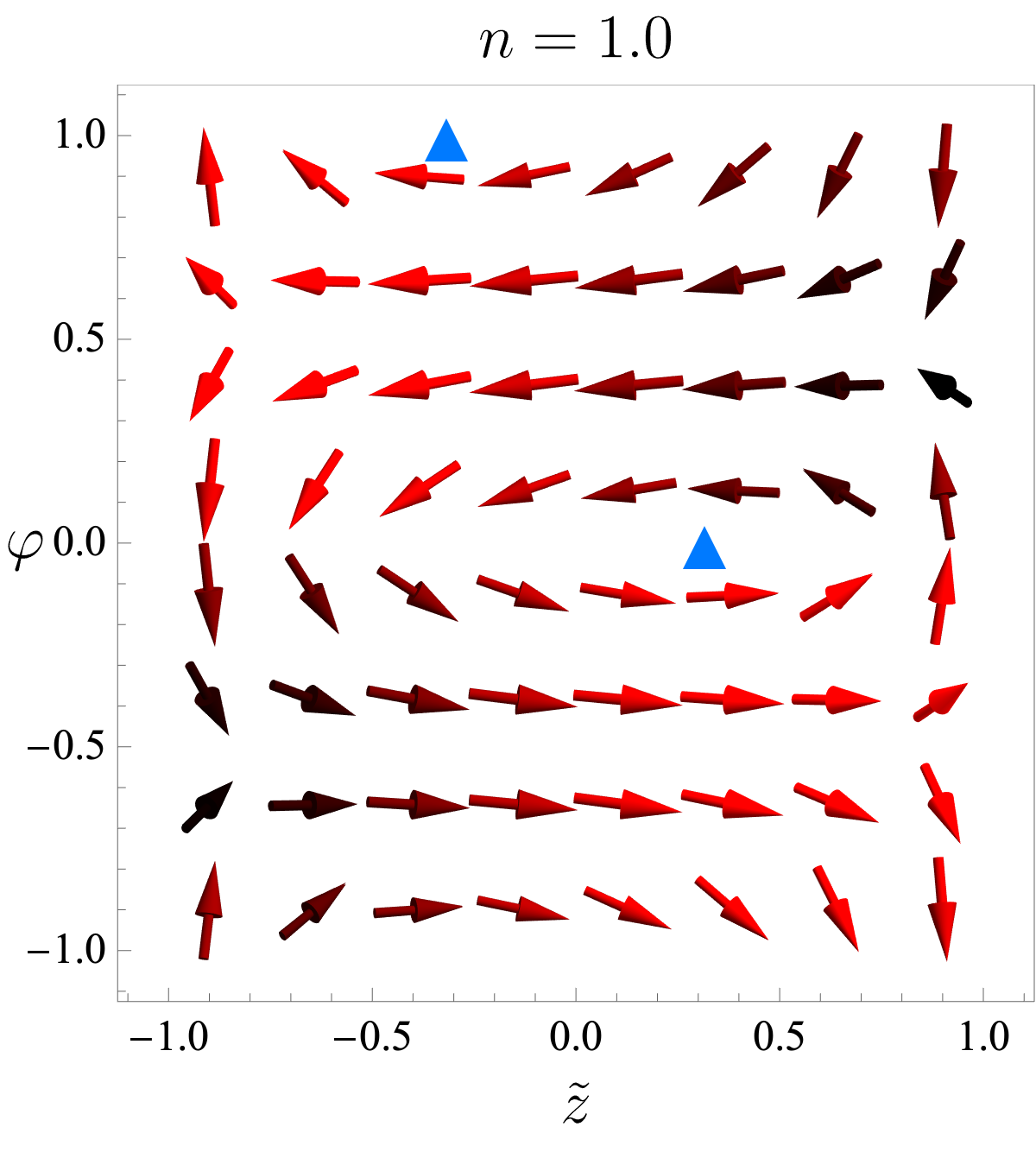}
	\includegraphics[height=.37\textwidth, trim={.6cm 0 0 0}, clip]{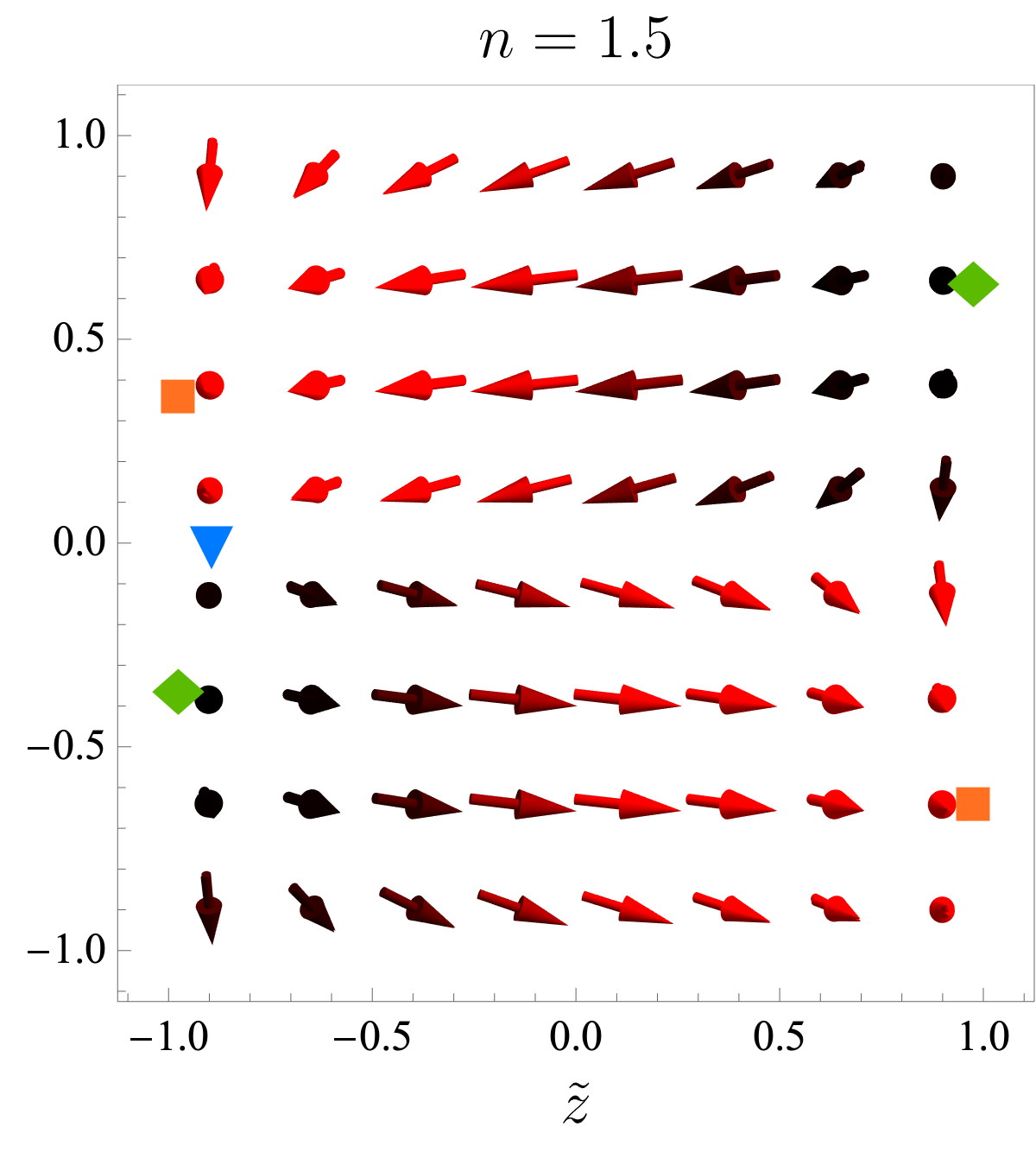}
	\includegraphics[height=.37\textwidth, trim={.6cm 0 0 0}, clip]{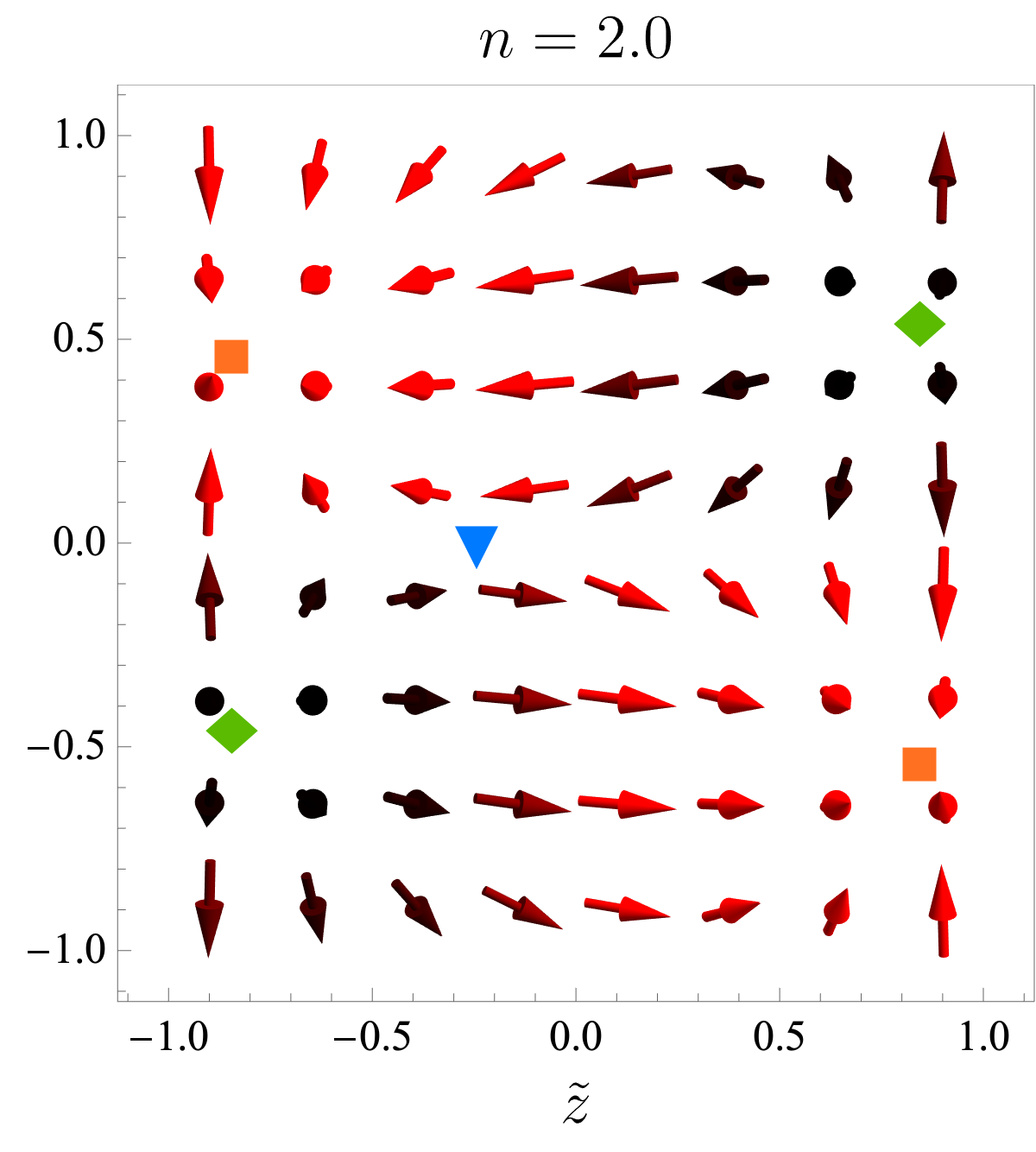} \\ %
	\caption{Plot of slices of the 3D parameter space for the nonlinear HN dimer model when $g = 0.7 J$ and $\Delta = 0.1 J$. \red{The panels represent bird-eye plots of different slices of constant $n$ values.} 
	The centers (blue, up triangles), saddles (blue, down triangles), stable (orange squares) and unstable (green diamonds) points, are those of the instantaneous HN dimer model when $gn\!\rightarrow\!\bar g$. The arrows represent the vector field described by Eqs.~\ref{eq:n dot} to \ref{eq:phi dot}, while their color is representative of the tilt along the $n$ direction: black indicates tilting in the negative direction, and red indicates the positive tilting.} 
	\label{fig:vector field}
\end{figure*}

Another helpful way to visualize this physics is shown in Fig.~\ref{fig:vector field}, where we plot arrows corresponding to the tangent vectors of dynamics on each point of the three-dimensional phase space.
\red{The panels} show the change in behavior of the fixed points on slices of fixed $n$: as long as for some time $t_1$, $gn(t_1) < g_\mathrm T$, the former are centers (\red{left panel}), whereas as soon as for some later time $t_2$ the condition $g n(t_2) > g_\mathrm T$ is met, the latter behave like saddles (see the middle and right panels).

In the intermediate regime, in the case of initial conditions close to $\varphi =0$, the system is pushed by the equations in a closed orbit around the center (top left panel in Fig.~\ref{fig:vector field}) and the population oscillates, as shown in Fig.~\ref{fig:aux1}. If the initial conditions are not sufficiently close to the fixed point of the blue up-triangle in the \red{left} panel, while moving around the center, the trajectory ends up close enough to the
basin of repulsion of unstable points (green diamonds) or to the basin of attraction of the stable points (orange squares).

\section{Conclusions and Outlook} \label{sec:conc}
In this paper, we have investigated both theoretically and experimentally the dynamics of two-site models with hopping asymmetry. In a linear model where the hopping asymmetry is externally fixed, we experimentally observed the transition from \pt-symmetric to \pt-broken regimes as the hopping asymmetry is increased. While all the orbits are closed in the \pt-symmetric regime, the population diverges in the \pt-broken regime due to the non-Hermiticity from the hopping asymmetry. In a nonlinear model where the hopping asymmetry is dynamically induced by population imbalance between the two sites, we experimentally observed three different regimes in behavior, depending on the initial coupling strength. In the weak and strong regimes, we observe stable population oscillations and exponential growth/decay of the population, respectively, similar to the behavior in the linear model described above. However, in the intermediate regime, we observe a coexistence of these dynamics, meaning that we can tune from stable oscillations to divergent behavior by simply varying the initial phase-difference between the two sites. As we explain, all three different regimes can be understood by studying the emergent fixed points of a closely-related nonlinear model in which the non-reciprocal hopping depends on the normalized population imbalance between the two sites. 

In the future, this work will pave the way towards the further exploration of non-Hermitian dynamics in more exotic systems. As demonstrated here and in our previous works~\cite{singhal2022measuring}, this mechanical platform can be used to simulate a wide-variety of lattice models, which would not be easy to realize in other systems. Going further, it will be interesting to explore, for example, the addition of other mean-field nonlinear effects~\cite{faugno2022interaction} as well as extensions to larger systems, \textit{e.g.}, such as non-Hermitian three-site trimer models~\cite{Franzosi2003chaotic,Liu2007Josephson,Lahaye2010Mesoscopic} or large lattices with many sites~\cite{molina1993dynamics}, where the interplay of gain and loss with artificial gauge fields and topological phenomena can also be explored~\cite{ozawa2019topological,keilmann2011statistically}. 

\section{Acknowledgments}

The experimental contributions (Y.~S., B.~G.) are based upon work supported by the National Science Foundation under grant No.~1945031. Y.~S. acknowledges support by the Philip J. and Betty M. Anthony Undergraduate Research Award and the Jeremiah D. Sullivan Undergraduate Research Award of the UIUC Department of Physics. T.~O. is supported by JSPS KAKENHI Grant No.~JP20H01845, JST PRESTO Grant No.~JPMJPR19L2, and JST CREST Grant No.~JPMJCR19T1. E.~M. and H.~M.~P. are supported by the Royal Society via grants UF160112, RGF\textbackslash EA\textbackslash 180121 and RGF\textbackslash R1\textbackslash 180071. This work was also supported by the BRIDGE Seed Fund for collaboration between the University of Birmingham and the University of Illinois at Urbana-Champaign.

\bibliographystyle{unsrt}
\normalem 

\end{document}